\documentclass[prb,reprint,doublecolumn,superscriptaddress, nobalancelastpage]{revtex4-1}

\usepackage{tikz}

\usepackage{etoolbox}
\patchcmd{\section}
  {\centering}
  {\raggedright}
  {}
  {}
\patchcmd{\subsection}
  {\centering}
  {\raggedright}
  {}
  {}

\usepackage{tcolorbox}
\usepackage[sort&compress]{natbib}
\usepackage{soul}
\usepackage{graphicx}
\usepackage{epstopdf}
\usepackage{verbatim}
\usepackage{float}
\usepackage{sidecap}
\sidecaptionvpos{figure}{c}
\usepackage{lipsum}
\usepackage{setspace}
\usepackage{dcolumn}
\usepackage{bm}
\usepackage{amssymb}
\usepackage{amsmath}	
\usepackage{mathtools}
\usepackage{color}
\usepackage{multirow}
\usepackage{lipsum}
\usepackage[a4paper]{geometry}
\newgeometry{top=1.25cm,right=1cm,left=1.75cm,bottom=1cm}
\usepackage[colorlinks=true, urlcolor=blue, pdfborder={0 0 0}]{hyperref}
\hypersetup{
linkcolor=blue,
citecolor=blue
}

\draft

\begin{document}

\title{\large{\textbf{Atomic-level Characterisation of Quantum Computer Arrays by Machine Learning}}}

\author{Muhammad Usman} \email{musman@unimelb.edu.au} \affiliation{Center for Quantum Computation and Communication Technology} \affiliation{School of Physics, University of Melbourne, Parkville, 3010, VIC, Australia.}

\author{Yi Zheng Wong} \affiliation{Center for Quantum Computation and Communication Technology} \affiliation{School of Physics, University of Melbourne, Parkville, 3010, VIC, Australia.}

\author{Charles D. Hill} \affiliation{School of Physics, University of Melbourne, Parkville, 3010, VIC, Australia.}

\author{Lloyd C.L. Hollenberg} \affiliation{Center for Quantum Computation and Communication Technology} \affiliation{School of Physics, University of Melbourne, Parkville, 3010, VIC, Australia.}
\maketitle
\onecolumngrid

\noindent
\textcolor{black}{
\normalsize{\textbf{Atomic level qubits in silicon are attractive candidates for large-scale quantum computing, however, their quantum properties and controllability are sensitive to details such as the number of donor atoms comprising a qubit and their precise location. This work combines machine learning techniques with million-atom simulations of scanning-tunnelling-microscope (STM) images of dopants to formulate a theoretical framework capable of determining the number of dopants at a particular qubit location and their positions with exact lattice-site precision. A convolutional neural network was trained on 100,000 simulated STM images, acquiring a characterisation fidelity (number and absolute donor positions) of above 98\% over a set of 17,600 test images including planar and blurring noise. The method established here will enable a high-precision post-fabrication characterisation of dopant qubits in silicon, with high-throughput potentially alleviating the requirements on the level of resource required for quantum-based characterisation, which may be otherwise a challenge in the context of large qubit arrays for universal quantum computing.}}}
\\
\twocolumngrid

\noindent
Of the leading platforms for the implementation of quantum computing architectures, qubits based on the spin of individual dopant atoms in silicon~\cite{Kane_Nature_1998, Loss_PRA_1998, Zwanenburg_RMP_2013, Fuechsle_NN_2012, Salfi_PRX_2018, Pla_Nature_2012, Morello_Nature_2010} are growing in interest given the nexus with nanoelectronics engineering and the long coherence times~\cite{Tyryshkin_Nat_Mat_2012, Saeedi_Science_2013}. For the exchange-based quantum computer design proposals~\cite{Kane_Nature_1998, Loss_PRA_1998, Pica_PRB_2016} where the physical separations between atomic qubits are small (10-15 nm), the pathway for scale-up to large two-dimensional arrays generally relies on uniformity of control of qubits and their interactions. Even small variations at the level of one lattice-site for qubits based on single or multiple dopant atoms can significantly affect the design and control of logical operations. While the details of few qubit systems can be determined using electrostatics and electron spin resonance~\cite{Wang_NatureSR_2016} and variations in interactrions mitigated by designing appropriate pulse schemes~\cite{Testlin_PRA_2007, Hill_PRL_2007}, for large-scale arrays a reliable and fast method of identification (atom count per qubit) and characterisation (exact spatial location of atoms in lattice) is critical. 

\noindent
Machine intelligence techniques have been extremely productive in a wide range of applications including material design, medical imaging, and data science, where the design space is enormously large~\cite{Butler_Nature_2018, Luna_Nature_2017, Libbrecht_NatureRG_2015, Murphy_NatureCB_2011} and/or autonomous predictions are required from big data analysis~\cite{Heureux_IEEE_2017}. In quantum devices, the application of deep learning for the automated fabrication of atomic-scale surface defects has been proposed~\cite{Rashidi_ACSN_2018, Rashidi_arXiv_2019}. This work integrates the high efficiency of machine learning algorithms towards pattern recognition~\cite{bishop2006pattern} with multi-million-atom simulations of Scanning-tunnelling microscope (STM) images of donor wave functions~\cite{Salfi_NatMat_2014,Usman_NN_2016} to formulate a theoretical framework with the capability of high-throughput and automated spatial metrology of the donor qubits in silicon. The ability to pinpoint the donor locations with exact-atom precision in large two-dimensional arrays will provide crucial input in the design and implementation of the fault-tolerant quantum computer architectures.     

\begin{figure*}
\includegraphics[scale=0.07]{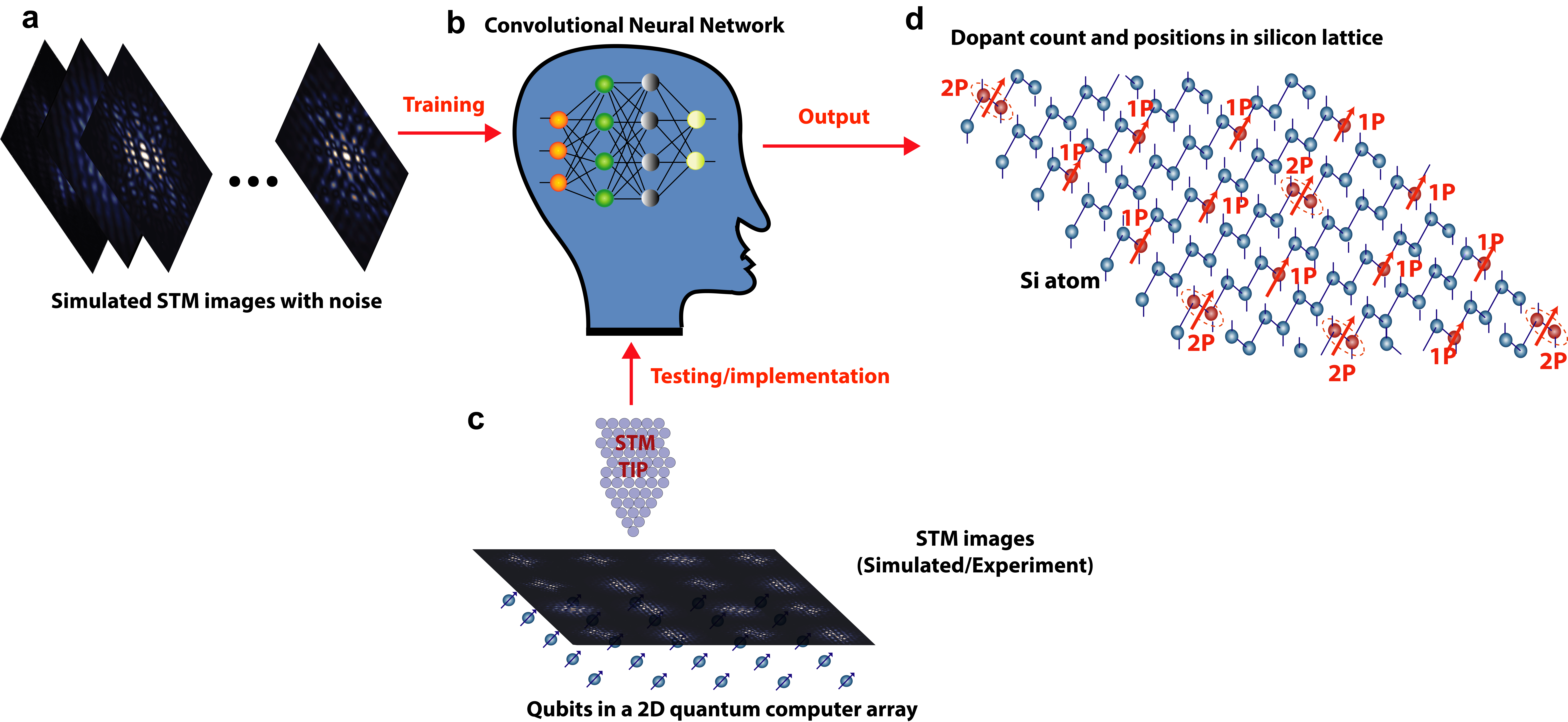}
\caption{\textbf{Overview of the automated atomic-level qubit characterisation technique:} \textbf{a.} A qubit is formed by electrons confined to either a single donor atom (1P) or a pair of closely-spaced atoms (2P) in silicon. Theoretically computed tunnelling current images of one electron wave functions confined on dopant qubits in silicon are generated. After including the application of noise typical of experimental images, the images are processed using an edge or feature detection analysis to reduce the computational and storage requirements. \textbf{b.} A large set (100,000) of the processed images is used to train a machine learning algorithm such as a convolutional neural network (CNN). \textbf{c.} The testing of the CNN is performed by generating a new set of 17,600 STM images with varying levels of planar and blurring noise. \textbf{d.} The trained CNN performs the exact-atom characterisation of qubits by precisely determining the spatial locations and count of dopant atoms corresponding to each test image.}
\label{fig:Fig1}
\end{figure*}
\noindent
STM has been extensively used to measure the spatially-resolved images of wave functions corresponding to the individual subsurface impurity atoms in various semiconductors, such as group V impurities in silicon~\cite{Salfi_NatMat_2014, Sinthiptharakoon_JPCM_2014}, Mn~\cite{Garleff_PRB_2008} and N~\cite{Ishida_Nanoscale_2015, Plantenga_PRB_2007} atoms in GaAs, and Bi atoms in InP~\cite{Krammel_PRM_2017}. Recently, STM imaging technique has been applied to determine the exact locations of single dopant atoms in silicon~\cite{Usman_NN_2016, Brazdova_PRB_2017}, which has opened new avenues to perform STM-based qubit characterisation and wave function benchmarking~\cite{Usman_Nanoscale_2017}. The idea underpinning the STM based dopant position metrology~\cite{Usman_NN_2016} was that the high resolution images of donor wave functions exhibit a map of features, in which the brightness and symmetry of the features directly encodes the information about the locations of atoms. A direct pixel-by-pixel comparison of a measured image with a library of theoretically computed STM images provided direct information about the exact dopant locations. This rigorous comparison approach worked-well for the individual atoms, but its scalability towards full-scale quantum computer arrays consisting of $O(10^6)$ qubits, where each qubit may consist of small clusters of closely-spaced donor atoms, is still an open problem and requires further development of computational techniques to efficiently process and characterise several thousand STM images. This work demonstrates that the high level of agreement between theory and experiment for these images at the pixel level opens the way for a machine learning approach based on training over large sets of simulated images, which could be implemented in future within an experimental set-up.

\noindent
Single-atom STM fabrication techniques~\cite{Fuechsle_NN_2012} can achieve the placement of phosphorus (P) atoms in silicon with accuracy in position of one lattice site, and the number of P atoms can in principle be controlled. However, the tunability of exchange interaction between a single P atom and two closely-spaced P atoms (2P) makes it an attractive qubit system~\cite{Wang_NQI_2016}, and the recent work has also studied qubits formed by three to four closely-spaced P atoms~\cite{Pakkiam_PRX_2018}. Therefore, the generalisation of the spatial metrology technique~\cite{Usman_NN_2016} beyond single donor atoms will broaden its scope for a wide range of qubit systems being considered for quantum computing applications. As the donor count per qubit increases, the number of available donor placement configurations drastically increase and impose stringent computational requirements for characterisation of qubits in large-scale devices. For example, merely increasing the number of dopant atoms per qubit from one to two leads to an increase in the possible position configurations from 60 to 1250 within 5 nm depth from the silicon surface. To enable an autonomous and robust spatial metrology of single donors and 2P dots in silicon, we perform the training of a convolutional neural network (CNN). The CNN learns the relationship between STM image features and the corresponding donor count and the exact spatial positions based on 10$^5$ simulated training images. The testing of the trained CNN over a large test data set consisting of 17600 simulated images including noise demonstrated a highly robust performance with fidelities above 98\% across the selected four depth planes. In principle, the donor atoms can be fabricated with a single target depth plane~\cite{Fuechsle_NN_2012}, in which case the qubit characterisation fidelities of 100\% are achievable from the established CNN framework.  

\begin{figure*}
\includegraphics[scale=0.28]{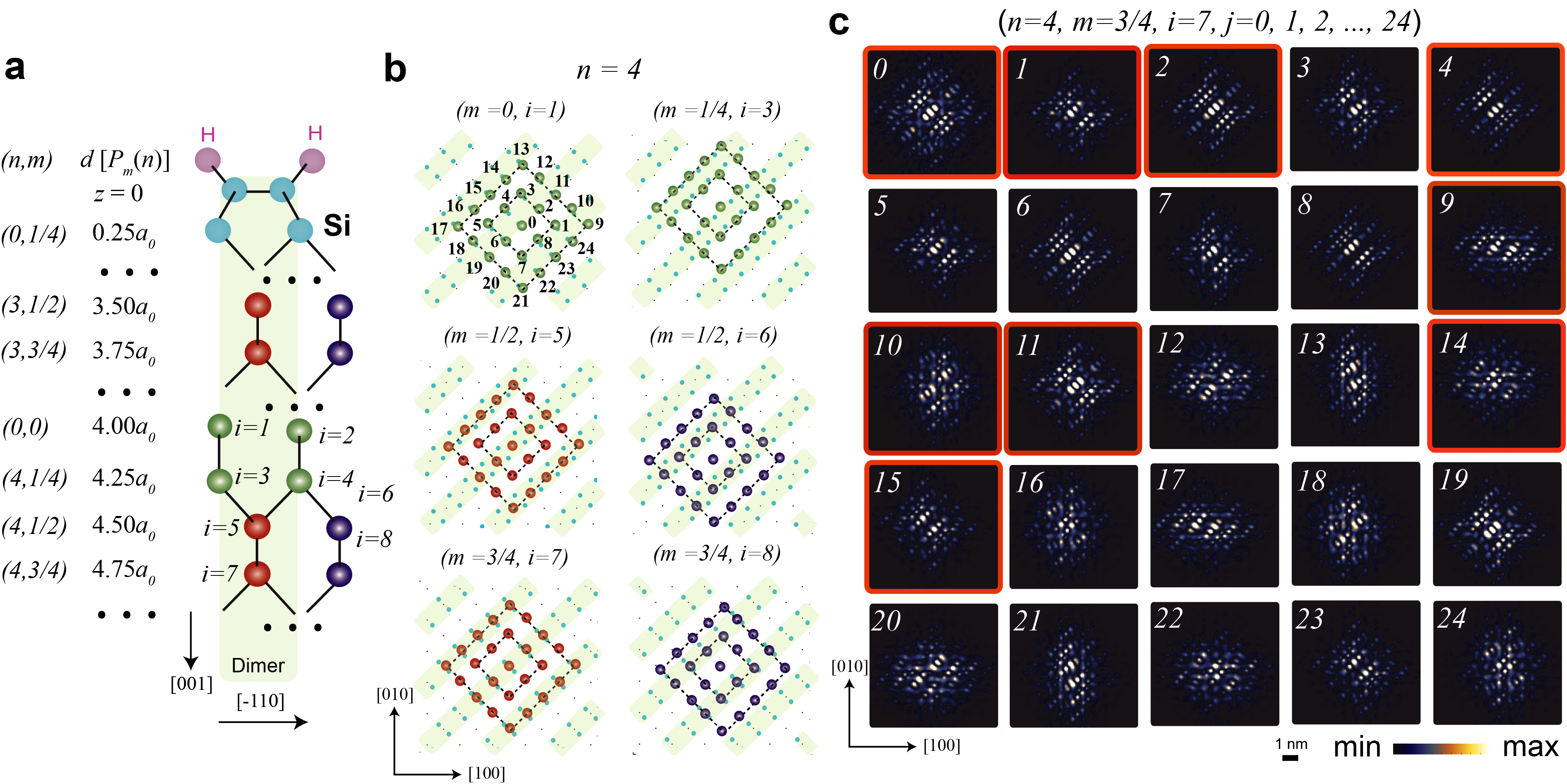}
\caption{\textbf{Symmetry analysis and classification of the computed STM images:} \textbf{a.} Schematic diagram of a small portion of the silicon lattice is shown, along with the positioning of the P donor atoms within a few nanometers of the $z$=0 surface. The $z$=0 surface is hydrogen passivated (purple atoms) and exhibits the formation of Si dimer rows (light blue atoms at $z$=0), which are aligned perpendicular to the page (along the [110] direction). The area is shaded underneath the dimers to provide guidance on the positioning of atomic sites with respect to the dimers. \textbf{b.} Based on symmetry of donor positions with respect to the location of surface dimer rows, six planes at $n$=4 are shown highlighting possible locations for donor atom placement. In each plane, the positioning of donor atoms is labelled by a number $j$, whose value varies from 0 to 24 as shown for $m=0$ and $i=1$ case. The position labels are same for the other five ($n,m$) cases. The central atom is marked as $j$=0 and the numbering in the inner ring is from 1 to 8 and in the outer ring is from 9 to 24 clockwise. \textbf{c.} Theoretically computed STM images are plotted for all possible positions ($j$=0, 1, 2, ..., 24) at $n$=4, $m$=3/4, and $i$= 7. The images clearly exhibit a well-defined symmetry of wave function features convoluted with the surface dimer positions. Based on the symmetry analysis, we find that the 2P images are identical when the second P atom is symmetrically distributed around the reference P atom at $j$=0. All distinct images are highlighted by a red colored boundary.}
\label{fig:Fig2}
\end{figure*}

\noindent
Figure~\ref{fig:Fig1} provides an overview of the proposed theoretical framework. To demonstrate the working of our technique, we have restricted each qubit formation to 1P and 2P configurations. The technique can be readily generalised to larger clusters consisting of a few closely-spaced P atoms per qubit. In part (a), one electron STM images are computed, where each image encodes the information about underpinning donor positions and count. In the next step, the computed STM images are processed via image reduction algorithms to increase the computational and storage efficiency of the machine learning framework. Two complementary methods are developed for image reduction, namely edge detection and feature averaging. Both methods drastically enhance the speed of the CNN. We also introduced various levels of planar and blurring noise to test the resilience of the trained CNN against realistic image distortions. Figure~\ref{fig:Fig1} (b) illustrates that the processed STM images are used to train a CNN. The testing of the trained CNN can be performed based on experimentally measured data and/or simulated STM images including noise as shown in Figure~\ref{fig:Fig1} (c). In this work, we have used simulated STM images with various levels of blurring and planar noise to test the performance of the CNN due to the unavailability of experimental data at present. The computation of the STM images have previously shown an unprecedented accuracy when compared directly with the STM measurements~\cite{Usman_NN_2016}, capturing both the symmetry and the brightness of the measured wave function features. Therefore, we expect that the training and testing of the machine learning framework performed in this study will be directly applicable to the experimental data sets available in the future. Figure~\ref{fig:Fig1} (d) shows the output of the CNN, indicating that it can accurately characterise each qubit by identifying donor count (1P or 2P) and their exact spatial locations in Si lattice. 
\\ \\
\noindent
\textcolor{blue}{
\textbf{\normalsize{Image classification and symmetry analysis}}} \\ \\
The STM images are computed by coupling the atomistic tight-binding calculations of the subsurface phosphorus dopant wave functions~\cite{Usman_JPCM_2015} with the Bardeen's tunnelling formalism~\cite{Bardeen_PRL_1961} and Chen's derivative rule~\cite{Chen_PRB_1990}. Note that we have performed this study for P in silicon system, however the developed machine learning framework can also be trained and applied to other group V donor atoms in silicon. In recent years, advancements in the atomic precision fabrication techniques~\cite{Fuechsle_NN_2012} have led to a donor atom placement accuracy to within $\pm$ $a_0$ in-plane variation for P donors in silicon, where $a_0$ is the silicon lattice constant. It was also shown that the donor atoms experience no diffusion along the growth direction (depth direction) when fabricated with a target depth of 4.75$a_0$~\cite{Usman_NN_2016}. In accordance with these published studies, we have assumed that the two closely-spaced dopant atoms in the case of 2P qubits are placed at the same depth from the Si surface. Furthermore, the distance between the two P atoms is within $2a_0$. We note that these are not limiting factors for our technique and robust qubit characterisation can be performed in the presence of donor depth variations and/or for larger donor separations.   

\begin{figure*}
\includegraphics[scale=0.24]{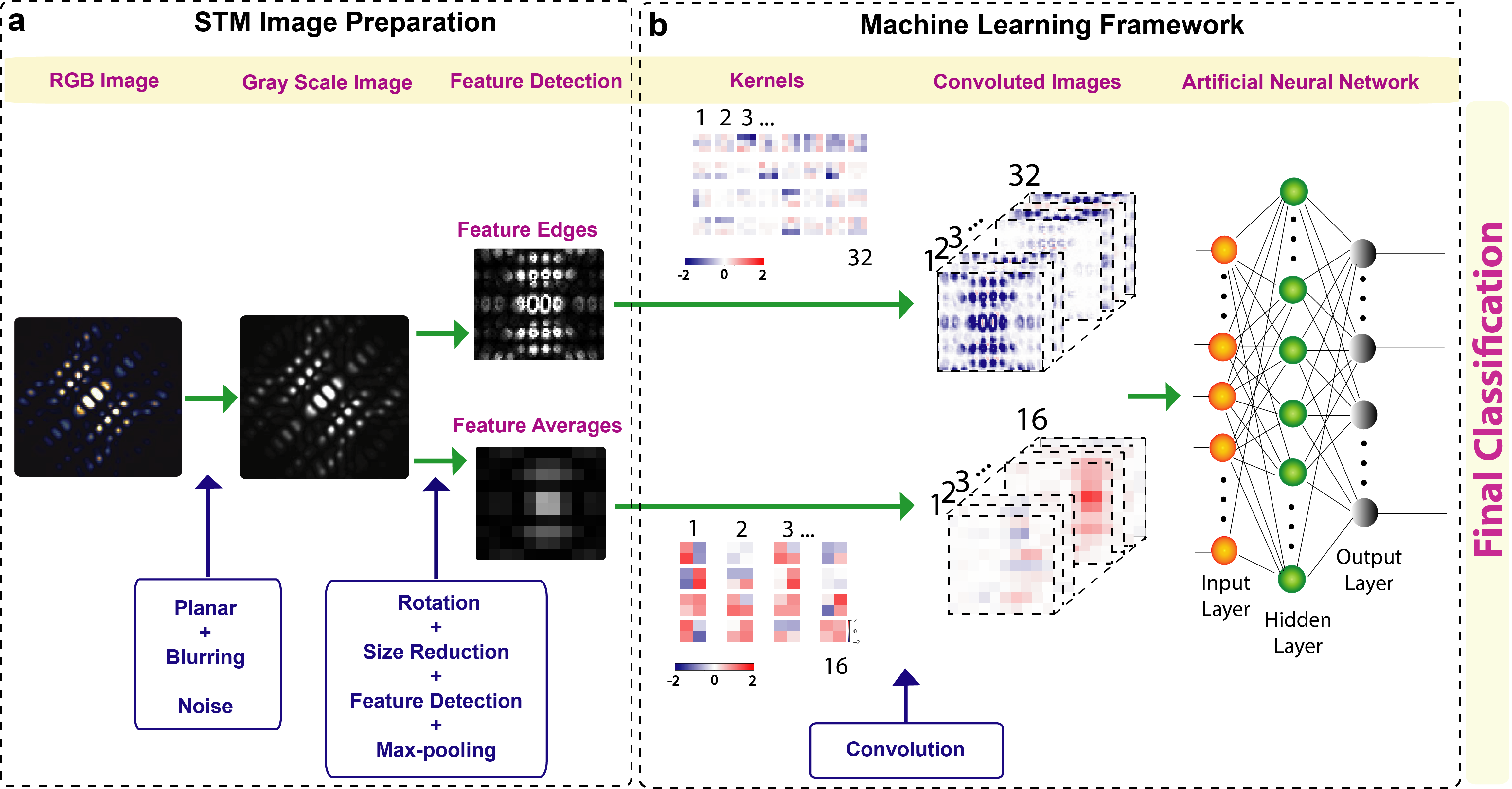}
\caption{\textbf{Flow chart diagram of machine learning framework:} \textbf{a.} For the demonstration of the working of our machine learning framework, we have selected one STM image corresponding to $n$=4, $m$=3/4, $i$=7, and $j$=2. The STM image is converted from RGB color plot to grayscale color plot to reduce the storage size. The STM image is further processed to extract either edges of the bright features or based on the average values over each bright feature (see the supplementary sections S3 and S4 for details). \textbf{b.} Kernels are shown with size 32 (3$\times$3) and 16 (2$\times$2) for the edge detection and the feature averaging schemes, respectively. Each training image is convoluted with the kernels to generate a set of 32 or 16 convoluted images. The convoluted images are used to train a neural network with one input layer, one hidden layer, and one output layer. The outcomes of the trained neural network classifies the STM images in accordance with the exact donor atom positions and count.}
\label{fig:Fig3}
\end{figure*}  

\noindent
A systematic labelling scheme was formulated to represent single donor atoms in silicon crystal~\cite{Usman_NN_2016}, which is extended in this work for a general case of qubits, where each qubit can be formed by either one donor atom (P) or two closely-spaced donor atoms (2P). Note that here 2P is defined as a donor cluster where two donor atoms are within the 2$a_0$ distance. The schematic diagram of a small portion of the Si crystal structure is shown in Figure~\ref{fig:Fig2} (a) to illustrate the possible locations for a dopant atoms to within a few nanometers from the $z$=0 surface. The $z$=0 surface is hydrogen passivated (shown by purple atoms and marked with H) and exhibits the formation of Si dimer rows (shown by light blue atoms), which are aligned perpendicular to the page (along the [110] direction). The area is shaded underneath the dimers to indicate the positioning of Si atomic sites with respect to the dimers. In our new notation, we represent each donor atom location by $L_m^{i,j}(n)$ and the corresponding STM images by ($n,m,i,j$), where $n$ selects a plane group, $m\in \{0, 1/4, 1/2, 3/4\}$ represents a plane within the group at depth $d[PG_m]$ = $(m+n)a_0$, $i$ identifies the positioning with respect to the surface dimer rows, and $j$ denotes the individual location(s) of the dopant atom(s) inside a selected plane defined by ($n,m,i$). Further details about this classification scheme are provided in the supplementary information section S1.

\noindent
For a given target depth based on ($n,m$), the dopant atoms are placed in the same plane. The in-plane positioning of the dopant atoms is shown in the Fig.~\ref{fig:Fig2}(b). To demonstrate the working of the machine learning framework, we have selected four target depths: 4$a_0$, 4.25$a_0$, 4.5$a_0$, 4.75$a_0$, corresponding to $n$=4 plane group. Due to the symmetry of the silicon crystal, the STM images exhibit same symmetry for other plane groups, therefore this particular set of planes at $n$=4 represents all types of STM images which repeat for other values of $n$~\cite{Usman_NN_2016}. We have separately plotted six planes corresponding to $n$=4. Note that for $m$=0 and 1/4, we have only plotted one value of $i$ (1 and 3). The positions corresponding to $i$ =2 and 4 are at the other edge of the dimer rows and symmetrically similar to the positions at $i$=1 and 3, respectively. This will result in exactly the same STM images, rotated by 270$^o$. The exact positions corresponding to these images can be determined by overlaying dimer row atoms~\cite{Usman_NN_2016}. In our classification scheme, we assume that one dopant atom is always at the center marked by $j$=0. The second donor in the case of 2P will occupy one of the locations at the boundaries of the two diamonds with distances $a_0$ and 2$a_0$ from the center dopant atom. These positions are labelled anti-clockwise from $j$=1 to 8 for the inner diamond and $j$=9 to 24 for the outer diamond as illustrated for ($n,m,i$)=($4,0,1$) in Figure~\ref{fig:Fig2} (b). Note that in each plane, the atom position at $j$=0 is same as the $i$ value in that plane.      

\noindent
Based on the dopant locations plotted in Figure.~\ref{fig:Fig2}(b), each dopant plane offers 25 possible configurations to place P/2P donor atoms, leading to 25 STM images. For the $n$=4 plane group, we computed in total 125 STM images. Figure~\ref{fig:Fig2} (c) plots the STM images for one selected plane corresponding to $m$=3/4 and $i$=7. Each STM image is labelled with the corresponding value of $j$. The STM images for the other five configurations are provided in the supplementary information Figures S1 to S5.  

\begin{figure*}
\includegraphics[scale=0.23]{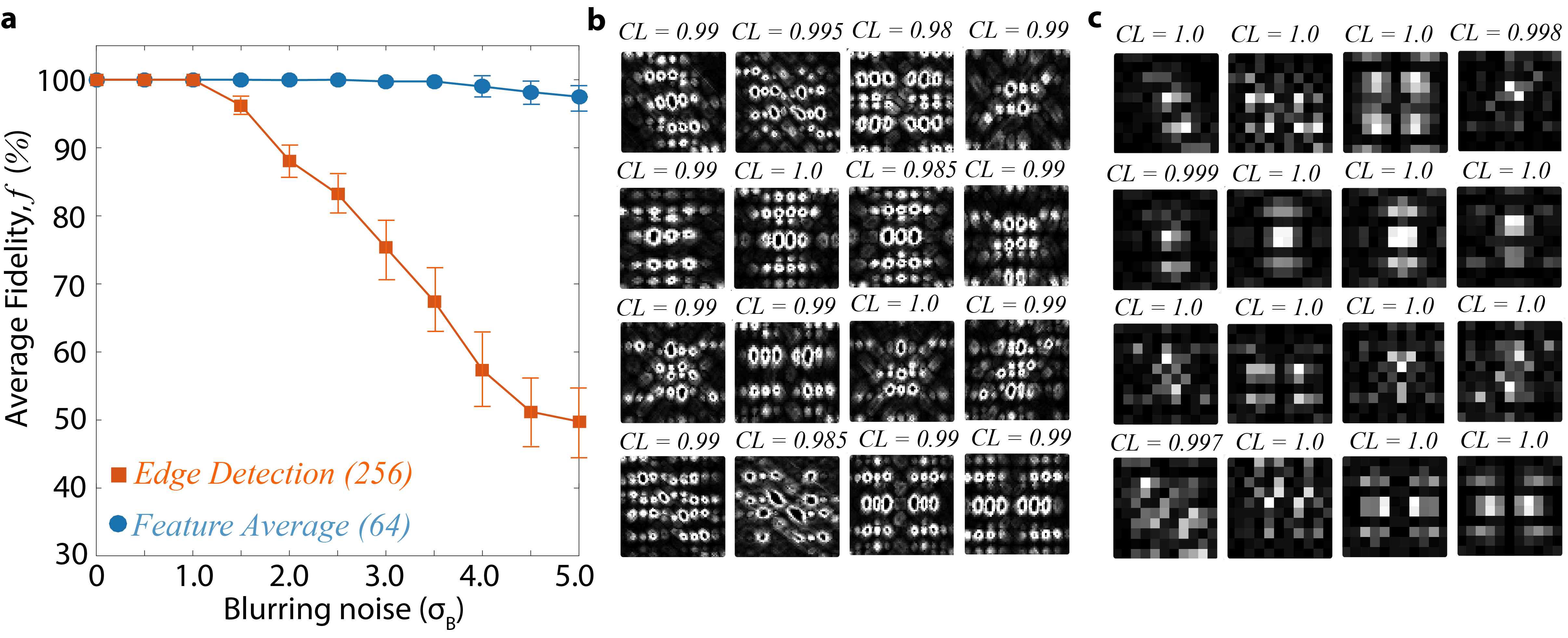}
\caption{\textbf{Test results from the machine learning framework:} \textbf{a.} The average fidelities from the CNN are plotted as a function of $\sigma_B$. At each value of $\sigma_B$, the average fidelity is computed based on 1600 test images (16 classes and 100 images per class). The error bars indicate the standard error of the mean based on two standard deviations. \textbf{b.} A set of 16 processed STM images are shown after the application of the edge detection procedure to test the working of the trained CNN. The images are applied random strengths of noise selected from 0 $\leq \sigma_P \leq$ 0.4 and 0 $\leq \sigma_B \leq$ 2.0 range. The corresponding unprocessed STM images are provided in the supplementary information Figure S15. In each case, the CNN correctly identifies the donor positions and count with the $CL$ values as provided on top of the images. \textbf{c.} A set of 16 processed STM images are shown after the application of the feature averaging procedure to test the working of the trained CNN. The images are applied random strengths of noise selected from 0 $\leq \sigma_P \leq$ 0.4 and 0 $\leq \sigma_B \leq$ 4.0 range. The corresponding unprocessed STM images are provided in the supplementary information Figure S15. In each case, the CNN correctly identifies the donor positions and count with the $CL$ values as provided on top of the images. }
\label{fig:Fig4}
\end{figure*} 

\noindent
From Figure~\ref{fig:Fig2} (b), we note that a number of dopant positions are equivalent due to their symmetrical distance from the center location at $j$=0. This implies that the corresponding STM images would also exhibit the same feature map with a possible rotation or reflection with respect to the axes parallel or normal to the dimer rows direction. For example, in Figure~\ref{fig:Fig2} (c), the images corresponding to $j$=4 and $j$=8 will be same if one of them is reflected with respect to the diagonal direction as shown in the supplementary information Figure S6. A careful examination of all images for ($4,3/4,7,j$) reveals that out of the 25 images, only 9 images are distinct. We classify the 25 images for the ($4,3/4,7,j$) group in 9 distinct image classes in the supplementary information Table T1. Further details about the classification of the STM images in distinct image classes is provided in the supplementary information section S1. Each class has been labelled by ($m,n,i,min(j)$), where $min(j)$ is the minimum value of $j$ in that class. For $n$=4, there are 50 distinct image classes. The 50 images representing the distinct classes are highlighted by the red color boundaries in Figure~\ref{fig:Fig2} (c) and also in the supplementary information Figures S1 to S5. The machine learning framework recognises dopant positions and count based on the feature maps, therefore it will only identify images with respect to these 50 classes. For example, in Figure~\ref{fig:Fig2} (c), the images corresponding to the positions $j$=$1,3,5,$ and 7 will be assigned to the same image class ($3/4,4,7,1$). The determination of the exact dopant locations within an image class can be subsequently performed based on its relative symmetry with respect to the positions of the surface dimer rows, which can be done by overlaying dimer atom positions on top of the image. 
\\ \\
\noindent
\textcolor{blue}{
\textbf{\normalsize{Application of noise and image size reduction}}} \\ \\
The computed STM images demonstrate a perfect symmetry and sharp bright features, whereas the published measured images~\cite{Brazdova_PRB_2017, Salfi_NatMat_2014, Usman_NN_2016} may consists of features which are asymmetrical in brightness and/or blurred around the edges. In order to test the resiliency of the machine learning framework in the presence of feature asymmetry and blurriness, we artificially apply a range of two types of noise to the computed images. A planar noise ($\sigma_P$) leads to an asymmetry of the features and a blurring noise ($\sigma_B$) causes the features to spread across their edges, making adjacent features harder to distinguish. The computation of noise and its application to the exemplary images is provided in the supplementary information section S2. The supplementary information Figures S8 and S12 plot computed STM images as a function of various strengths of the planar and blurring noise, respectively. Based on the plotted images, we infer that $\sigma_P \leq$ 0.4 and $\sigma_B \leq$ 4.0 are the reasonable range of noise strengths beyond which the computed STM images become significantly distorted and cannot be accurately recognised. As part of the STM image preparation process, the application of noise is performed in the second step as illustrated in Figure~\ref{fig:Fig3} (a).

\noindent
 After the addition of noise, the computed STM images are further processed to reduce their size. The size of a computed image is 535$\times$535 pixels, which is quite large for the purpose of training and testing of a machine learning framework, which generally requires processing of several thousand images (10$^5$ training and 17600 test images in our study). To reduce the computational burden, we apply image reduction steps. Each coloured pixel represented by the RGB format is first converted to the grayscale format. We note that the STM images are computed over a large area (8$\times$8 nm$^2$), however the area around the features is dark indicating negligible tunnelling current. As the information about the donor positions is encoded in the bright features, we crop the dark region to further reduce the image sizes. This is done by first rotating the image clockwise by 45$^o$ degrees and then removing the pixels with the tunnelling current values below a threshold value. Further details of this process are provided in the supplementary information section S3 and S4, along with the supplementary Figures S13 and S14. At the end of this process, the image size is reduced from 535$\times$535 pixels to about 237$\times$189 pixels.

\noindent
The information about the donor positions is present in the size, arrangement, and brightness of the image features. It is noted that each image consists of twenty to thirty bright features, which distinctly describe the corresponding position(s) and count of dopant atom(s). To further reduce the size of an image, we focus on the bright features and apply two techniques to extract the feature properties while preserving the donor position information. The first method which focuses on the shape of the features is called feature edges scheme in Figure~\ref{fig:Fig3} (a). Further details about this method are described in the supplementary information section S3. In this technique, we apply a filter operation which extracts the edges of the features. The image is then 3$\times$3 sub-sampled or max-pooled to obtain a final image consisting of about 79$\times$63 pixels. Note that in this method, each image is of slightly different size based on number of features and their spatial distributions. Overall, the size of the computed images after the edge detection processing is always below 90$\times$90 for all the images studied in this work. 

\noindent
The second method is labelled as feature averages in Figure~\ref{fig:Fig3} (a) and is described in detail in the supplementary information section S4. In this scheme, we represent each feature by its overall average brightness with respect to the dimer positions. This drastically reduces the size of an image to 11$\times$10 pixels. Moreover, the size of the final processed images is also fixed for all cases. We do not apply any max-pooling function to feature averaged images. Following the image processing steps, we train and test a machine learning framework for both methods separately. A comparison is performed between the two image reduction schemes based on the computational efficiency and robustness against the application of noise. 
\\ \\
\noindent
\textcolor{blue}{
\textbf{\normalsize{Training of the convolutional neural network}}} \\ \\
The processed STM images are used to train a convolutional neural network (CNN). The robust training of a CNN generally requires a very large data set, typically consisting of sample spaces with $O(10^3)$ sizes. We used ideal images and the images with various levels of planar noise ($\sigma_P$) to train the CNN. To construct a sufficiently large training data set, we randomly vary $\sigma_P$ between 0 and 0.4 and compute 2000 images corresponding to each of the 50 classes, accumulating a library of 10$^5$ training images. These images are separately processed through the edge detection and feature averaging schemes and are used to train two independent CNNs with one input, one hidden and one output layers. 

\noindent
Figure~\ref{fig:Fig3} (b) displays the work-flow of the CNN for the established high-throughput qubit characterisation scheme. Each image in the training data set is passed through the convolution layer before setting-up the CNN. In the case of the edge detection scheme, the CNN consists of a convolutional layer with 32, 3$\times$3 kernels along with 2$\times$2 max-pooling, followed by a hidden layer of 256 rectified linear units (ReLu) activated neurons. The images are scaled to 48$\times$48 pixels. Training on 10$^5$ images with 30 epochs achieved a learning accuracy of above 99.5\% and completed in about 5 hours on an average desktop machine. For the case of the feature averaging scheme, a hidden layer of 64 ReLu activated neurons, and the training was performed on 10$^5$ images with 20 epochs, which was completed in about 30 minutes on an average desktop machine and resulted in a learning accuracy of 100\%. In both cases, the output layer is densely connected layer with Softmax activation function. The CNN was compiled based on the Adam algorithm~\cite{Kingma_arxiv_2017} with the learning rate of 10$^{−4}$ and the categorical cross-entropy for optimization and as the loss function, respectively. The number of neurons is optimised by testing out various configurations of the CNN, and a sufficiently low number of neurons that will maintain the near perfect learning is chosen. The implementation of the CNN was performed by using Keras~\cite{chollet2015keras}, utilizing TensorFlow as the underlying platform~\cite{tensorflow2015-whitepaper}.      
\\ \\
\noindent
\textcolor{blue}{
\textbf{\normalsize{Qubit characterisation fidelities including noise}}} \\ \\
To test the performance of the machine learning framework, we define two parameters as the fidelity ($f$) of the qubit characterisation and the confidence level ($CL$). For a given test image, the trained CNN returns a set of 50 values (between 0 and 1), where each value indicates $CL$ for that image to be in one of the 50 image classes. The test image is characterised as belonging to a particular image class based on the highest $CL$ value. If the highest $CL$ correctly identifies the image class, it is assigned a value of $f$=1, otherwise $f$=0. To test the robustness of the CNN, we prepared three separate test sets for both the edge detection and the feature averaging schemes. The first test set consists of 50 ideal STM images without the application of noise and the trained machine learning framework resulted in $f$=1 with $CL$=1 for all images. This confirmed that the CNN has been properly trained based on the prepared training images.       

\noindent
The second case consisted of test images after the application of blurring noise only for both the edge detection and the feature averaging schemes. To establish a sufficiently large test set, we arbitrarily selected 16 STM image classes (see supplementary information section S5 and Figure S15 for details) and applied the blurring noise ($\sigma_B$) with its strength varying from 0 to 5.0 pixels with an increment of 0.5. At each value of $\sigma_B$, its orientation is randomly varied and 100 image are computed. The total test set consisted of 17600 STM images from the 16 classes. In the supplementary information Figures S16 and S17, we have plotted the \% fidelity values (number of correctly classified images out of the 100 noisy images) obtained from the CNN for each image class independently. Figure~\ref{fig:Fig4} (a) plots the average values computed from the 1600 images (16 classes $\times$ 100 noise orientations) at each value of the $\sigma_B$. The error bars indicate the standard error of the mean based on two standard deviations of the data. As expected, fidelities decrease when $\sigma_B$ increases and the images become harder to recognise. Based on the plotted results, we infer that the feature averaging scheme provides much higher fidelities compared to the edge detection scheme for large values of $\sigma_B$. The higher fidelity values for the feature averaging scheme are also coupled with about an order of magnitude better computational efficiency and two orders of magnitude lower storage requirements. Therefore, we conclude that the feature averaging scheme offers superior performance for the established machine learning based qubit characterisation compared to the edge detection scheme. Interestingly, we find that the fidelity drop varies between different image classes and some images offer very high resiliency against the application of $\sigma_B$. This information may provide a useful input for the selection of a target depth during donor atom fabrication processes incorporating this autonomous characterisation scheme.

\noindent
In the final test set, we simultaneously apply both planar and blurring noise to the set of STM images plotted in the supplementary information Figure S15. In the case of the edge detection scheme, we randomly vary noise orientation and strength from 0 $\leq \sigma_P \leq$ 0.4 and 0 $\leq \sigma_B \leq$ 2.0 range, whereas for the feature average scheme, we randomly vary noise levels from 0 $\leq \sigma_P \leq$ 0.4 and 0 $\leq \sigma_B \leq$ 4.0 due to its higher resiliency against the application of $\sigma_B$. The final processed images including noise are shown in Figure~\ref{fig:Fig4} (b) and (c). For both image-reduction schemes, the CNN characterises each image correctly ($f$=1), with the $CL$ values shown in the figure. Based on these results, we conclude that the CNN has been trained to accurately identify the STM images in the presence of both planar and blurring noise.  
\\ \\
\noindent
\textcolor{blue}{
\textbf{\normalsize{Summary and Outlook}}} \\ \\
\noindent
In summary, this work takes a first step towards implementation of a machine learning framework for autonomous characterisation of a large-scale quantum computer architectures based on dopant impurities in silicon. The input to the established framework are simulated STM images of one electron wave functions confined on single dopants or on small clusters of closely-spaced dopants. The images are processed to optimise the exploitation of information known about the system (\textit{e.g.} lattice geometry and surface dimers) and to reduce computational burden by developing and applying two feature-detection methods, namely edge detection and feature averages. Our results showed that both feature-detection methods enable high fidelity qubit characterisation at low noise level, with the feature averaging method providing considerably superior performance in the presence of large blurring noise. A convolutional neural network (CNN) is trained to characterise the noisy STM images and pinpoint the corresponding dopant atom position(s) and count with an exact lattice site precision. For the purpose of demonstrating the working of the established methods, the CNN was trained and tested on simulated STM images including noise levels commensurate with the published measurements. We note that the computed STM images have previously shown an extremely good agreement with the measured images at both pixel-by-pixel and feature-by-feature levels~\cite{Usman_NN_2016}, therefore we expect that the trained CNN will be able to characterise experimental images with an accuracy equivalent to the simulated images with noise. As the training of CNN requires several thousand images, the capability to train based on simulated images eliminates the need for performing large scale experimental measurement, saving a lot of time and effort.     

\noindent
The established automated characterisation of atomic qubits with such a high level of accuracy will assist in the design and implementation of two-qubit quantum gates. The underpinning experimental expertise, the atomic-precision fabrication of dopant atoms in silicon via STM lithography~\cite{Fuechsle_NN_2012} and the STM images of dopant wave functions by low-temperature tunnelling of single electron~\cite{Salfi_NatMat_2014}, has already been demonstrated. Augmentation of the formulated machine learning setup with these experimental techniques is expected to enable high-throughput characterisation post-fabrication with minimal human interaction. We envision that as the number of qubits in quantum devices grows, the characterisation by direct quantum measurements will be increasingly onerous, and a fast, reliable and autonomous methodology may play a crucial role in the scale-up process. 

\noindent
\\
\textcolor{blue}{
\textbf{\normalsize{Methods}}} \\ \\
\noindent
\textbf{\normalsize{Tight-binding wave function calculations}} \\
The computation of phosphorus dopant wave functions is performed by solving an atomistic sp$^3$d$^5$s$^*$ tight-binding Hamiltonian~\cite{Boykin_PRB_2004}. The P donor atom is placed in a large silicon box consisting of roughly four million atoms. The confining potential on the P atom is represented by a comprehensive description of the central-cell effects, which include non-static dielectric screening of donor potential~\cite{Usman_JPCM_2015}:

\begin{equation}
	\label{eq:Nonstatic_donor_potential}
	U \left( r \right) = \frac{-e^2}{ \epsilon \left( 0 \right) r} \left( 1 + A \epsilon \left( 0 \right) \mathrm{e}^{- \alpha r} + \left( 1-A \right) \epsilon \left( 0 \right) \mathrm{e}^{- \beta r} - \mathrm{e}^{- \gamma r}  \right)
\end{equation} 

\noindent
where A, $\alpha$, $\beta$, and $\gamma$ are fitting constants and have been numerically fitted as described in the literature~\cite{Nara_JPSJ_1965}. Additionally, the nearest-neighbor bond-lengths of Si:P are strained by 1.9\% in accordance with the recent DFT study~\cite{Overhof_PRL_2004}. The value of U$_0$ at the donor atom site is adjusted to empirically fit the binding energies of 1s manifold of states~\cite{Usman_PRB_2015}. The calculation of wave functions also included the effect of 2$\times$1 surface reconstruction, leading to the formation of dimer rows at the $z$=0 surface~\cite{Usman_NN_2016, Craig_SS_1990}. The impact of the surface strain due to the 2$\times$1 reconstruction is included in the tight-binding Hamiltonian by a generalization of the Harrison's scaling law, where the inter-atomic interaction energies are modified with the strained bond length $d$ as $(\frac{d_0}{d})^{\eta}$, where $d_0$ is the unperturbed bond-length of Si lattice and $\eta$ is a scaling parameter whose magnitude depends on the type of the interaction being considered and is fitted to obtain hydrostatic deformation potentials~\cite{Boykin_PRB_2004}. The boundary conditions for the silicon box are selected as closed, with dangling bond energies shifted by large values to exclude their effect in the working range of energy~\cite{Lee_PRB_2004}. The theoretical calculations were performed using the NEMO-3D framework~\cite{Klimeck_2, Ahmed_Enc_2009}.  
\\ \\
\noindent
\textbf{\normalsize{Computation of STM Images}} \\
The computation of the STM images is performed by coupling the atomistic tight-binding wave function calculation with the Bardeen's tunnelling current formalism~\cite{Bardeen_PRL_1961}. The wave function is decayed in the vacuum region above the reconstructed silicon surface based on the Slater orbital real-space representation~\cite{Slater_PR_1954}. For the calculation of the tunnelling current, the dominant contribution has been found to come from the $d_{z^2 - \frac{1}{3}r^2}$ tip orbital~\cite{Usman_NN_2016}, which is computed by applying the derivative rule reported by Chen~\cite{Chen_PRB_1990}:

\begin{equation}\label{func}
{\textrm I_\textrm {T}} (r_0) \varpropto \left\lvert \frac{2}{3}\frac{\partial^2 \Psi_ \textrm D (r)}{\partial z^2} - \frac{1}{3}\frac{\partial^2 \Psi_ \textrm D (r)}{\partial y^2} - \frac{1}{3}\frac{\partial^2 \Psi_ \textrm D (r)}{\partial x^{2} } \right\rvert _{r_0} ^2
\end{equation}

\noindent where $\Psi_{\rm D}$ is the donor wave function and $r_0$ is the position of the STM tip.
 
\noindent
Each computed STM image is spanned over 8$\times$8 nm$^2$ area and consists of about 535$\times$535 pixels represented in the RGB color scheme. 

\noindent
\\
\textbf{Acknowledgements:} This work was supported by the Australian Research Council (ARC) funded Center for Quantum Computation and Communication Technology (CE170100012), and partially funded by the USA Army Research Office (W911NF-08-1-0527). Computational resources were provided by the National Computing Infrastructure (NCI) and Pawsey Supercomputing Center through National Computational Merit Allocation Scheme (NCMAS). 
\\ \\
\noindent
\textbf{Author contributions:} LCLH conceived the initial idea, subsequently developed by all authors. MU and LCLH planned and supervised the project. MU and YZW computed and processed the STM images with input from LCLH. YZW, MU, and CDH formulated and applied the machine learning framework. All authors contributed in the discussions and analysis of the data. MU wrote the manuscript with inputs from all authors.
\\ \\
\noindent
\textbf{Data availability:}
The data that support the findings of this study are available within the article and its supplementary information file. Further requests can be made to the corresponding author. 
\\ \\
\noindent
\textcolor{black}{
\textbf{Additional information:}} The accompanied supplementary information document consists of 5 sections, 6 tables, and 17 figures. It provides additional STM images, a description of the feature-detection schemes, the implementation and application of noise, and benchmarking of the machine learning framework.   
\\ \\
\noindent
\textbf{Competing financial interests:} The authors declare no competing financial interests. A provisional patent application has been submitted based on aspects of this work. 

%\end{document}

\clearpage
\newpage

\noindent
\large{\textbf{\underline{Supplementary Information Document}}}

%\newbibliography{SM}

\noindent
\\ \\
%\title{High-throughput exact-atom qubit metrology by combining STM imaging with machine learning}
\title{\large{\textbf{Atomic-level Characterisation of Quantum Computer Arrays by Machine Learning}}}
%\title{\large{Towards autonomous SPM fabrication of quantum computer array by machine learning }}

\normalsize
\author{Muhammad Usman} \email{musman@unimelb.edu.au} \affiliation{Center for Quantum Computation and Communication Technology, School of Physics, University of Melbourne, Parkville, 3010, VIC, Australia.}

\author{Yi Z. Wong} \affiliation{Center for Quantum Computation and Communication Technology, School of Physics, University of Melbourne, Parkville, 3010, VIC, Australia.}

\author{Charles D. Hill} \affiliation{School of Physics, University of Melbourne, Parkville, 3010, VIC, Australia.}

\author{Lloyd C.L. Hollenberg} \affiliation{Center for Quantum Computation and Communication Technology, School of Physics, University of Melbourne, Parkville, 3010, VIC, Australia.} 
\onecolumngrid

\normalsize

\noindent
\section*{S1. Classification and symmetry analysis of STM images}

\noindent
Figure 2 (a) in the main text shows a small portion of the silicon crystal along with the atom position labels. Along the [001] direction, the Si lattice planes are divided into four plane groups ($PG_m$, where $m \in$ \{0, 1/4, 1/2, 3/4\}). Each plane group is a set of planes $P_{m}(n)$, whose depths from the $z$=0 surface are given by: $d[P_{m}(n)]$=($m$+$n$)$a_0$, where $n$=0,1,2,3,... Note that $(m,n)$=(0,0) represents the $z$=0 surface. Considering the periodicity of the dimers along the [-110] direction, two possible dopant locations $L_{m}^i(n)$ per plane $P_{m}(n)$ are defined by $i=8m+1$ and $i=8m+2$, which repeat periodically in the lateral direction. The positions within the $L_{1/2}^i(n)$ and $L_{3/4}^i(n)$ groups are either directly underneath the dimers marked with green color or in-between the two dimers marked with red color. As the lateral positions within the $L_{0}^i(n)$ and $L_{1/4}^i(n)$ groups are always at the edges of the dimers, so these are equivalent with respect to the surface strain and therefore are marked by a single blue color. At $n$=4, we highlight the atomic positions labelled with the corresponding $i$ values. 
\\ \\
\noindent
For the purpose of this study, we have considered qubits formed by either single phosphorus (P) atoms or a pair of P atoms that are closely-spaced as shown in the Figure 2 (a) and (b) of the main manuscript. The target depths are selected as 4$a_0$, 4.25$a_0$, 4.5$a_0$, and 4.75$a_0$ from the reconstructed silicon surface (corresponding to $n$=4 group), where $a_0$ is the silicon lattice constant (0.543095 nm). Based on the symmetry of donor positions with respect to the surface dimer rows, the available locations for the P donor atoms are shown in Figure 2 (b) of the main manuscript. Each donor position is uniquely classified by 4 numbers ($n, m, i, j$), where $n$ is the target plane depth group (i.e. 4 for this study), $m$ is the plane identifier ($0, 1/4, 1/2, 3/4$), $i$ is the atom number inside the group (1, 2, $\cdots$, 8) along the depth direction ([001]), and $j$ is the actual location of the P atoms for given ($n, m, i$) and varies from 0 to 24. The $j$=0 label corresponds to a single donor atom (P) and $j >$ 0 indicates two phosphorus atoms (2P), where the position of one P atom is always fixed at $j$=0 and the second P atom is placed in one of the neighboring positions marked by $j$=1,2,3, $\cdots$, 24. We should point out that this configuration, where qubits are formed by single donors or pairs of donors, is an exemplary case to demonstrate the working of our automated metrology. The technique is applicable for a large number of configurations, for example where donor clusters are made up of more than 2 closely spaced atoms, or where the donor positions within the cluster may vary along the [001] direction.      
\\ \\
\noindent
Figure 2 (c) in the main text shows all the STM images computed for $j$=0,1,2,3, $\cdots$, 24 at ($n, m, i$) = ($4, 3/4, 7$). The computed STM images for ($n, m, i$) equal to ($4, 0, 1$), ($4, 1/4, 3$), ($4, 1/2, 5$), ($4, 1/2, 6$) and ($4, 3/4, 8$) are plotted in the Supplementary Figures~\ref{fig:FigS1} to~\ref{fig:FigS5}, respectively. 
\\ \\
\noindent
Based on the symmetry of the silicon crystal, the atomic locations for phosphorus atoms are also symmetric with respect to the center position ($j$=0). This leads to many of the STM images exhibiting the same symmetry and brightness of features, with rotation and/or reflection with respect to the axes parallel or perpendicular to the surface dimer rows. As an example, in the Supplementary Figure~\ref{fig:FigS6} (a), we have plotted an STM image for the donor position at ($4, 3/4, 7, 4$) along with a green dotted line indicating the diagonal axis through the center of the image parallel to the dimer row direction. When this image is reflected along the green axis in part (b), it corresponds to the donor position at ($4, 3/4, 7, 8$). Therefore, we conclude that these two donor positions (($4, 3/4, 7, 4$) and ($4, 3/4, 7, 8$)) are identical with respect to the silicon crystal symmetry and lead to STM images with a similar feature distribution. In the Supplementary Figures~\ref{fig:FigS1} to~\ref{fig:FigS5}, we have highlighted unique STM images by plotting red color boundaries and the other STM images are symmetry identical of these images. The machine learning framework will identify only one donor position for all the STM images that are symmetry identical. We can group such positions into a same image symmetry class. By carefully analyzing the symmetries of all the computed STM images, we have grouped them in the classes, where each class consists of images with same feature maps but within rotation and/or reflection operation. Supplementary Tables~\ref{tableS1} to~\ref{tables6} provide classification of the STM images in unique image classes and also describe the relationship (symmetry, rotation and/or reflection) between the images within the same class. Based on this classification, we have identified in total fifty classes of unique STM images. The machine learning algorithm will be trained based on these fifty image classes. After training process, a test image will be identified based on a probability distribution consisting of fifty numbers, where each number represents the probability or confidence level ($CL$) for a particular test image being in one of the fifty classes. The test image will be recognized and characterized based on the highest probability value ($f$=1).

\section*{S2. Application of Noise}

\noindent 
The theoretically computed STM images show bright features that are sharp and symmetrically distributed across or along the dimer row axes. However, the measured images reported in the literature~\cite{Salfi_NatMat_2014, Usman_NN_2016, Brazdova_PRB_2016, Sinthiptharakoon_JPCM_2014} may exhibit features which are slightly asymmetrical and/or blurred close to the edges. To test the working of our machine learning technique in the presence of feature asymmetry and blurriness, we introduce artificial noise in the simulated images. Below, we briefly describe two methods that have been used in this work for the implementation and application of noise. 

\subsection*{Planar Noise}

\noindent
The effect of asymmetry in the brightness of image features is captured by simulating a planar noise which is based on computing a two-dimensional function as follows: 

\begin{equation}
	\label{eq:Nonstatic_donor_potential}
	N(x,y) = 1 + N_z + N_x(x-x_0) + N_y(y-y_0)
\end{equation} 
\noindent
Where $N_x$, $N_y$ and $N_z$ are independently and randomly generated from a Gaussian distribution with $\mu$=0 and $\sigma_P$ set to various noise levels. Here, $x_0$ and $y_0$ are arbitrarily chosen to be close to the center of the image. A random sample of the contours of such a plane created by $N(x,y)$ is shown in Supplementary Figure~\ref{fig:FigS13}. 
\\ \\
\noindent
The resulting STM image including the planar noise can be simulated by using:

\begin{equation}
	\label{eq:Nonstatic_donor_potential}
	I_{noise}(x,y) = I_{ideal}(x,y) \times N(x,y)
\end{equation} 
\noindent
Where $I_{noise}(x,y)$ and $I_{ideal}(x,y)$ are the computed STM images with and without the application of planar noise, respectively. The STM images for position ($4,3/4,7,0$) with various levels of the planar noise ($\sigma_P$) are shown in the Supplementary Figure~\ref{fig:FigS14}. The plotted images show that for $\sigma_P \geq$ 0.5, the images are severely cropped and hence do not represent a realistic condition. We therefore infer that a value of $\sigma_P$ between 0 and 0.4 is reasonable to test the resiliency of our machine learning framework against the presence of planar noise. 
\\ \\
\noindent 
The training set for the neural network requires a large number of STM images, therefore we included the planar noise in the training set by varying its magnitude randomly between 0.01 and 0.4. To create a reasonably large training set, we computed 2000 images per image class, producing a test set of 10$^5$ image for all of the 50 classes.

\subsection*{Blurring Noise}

\noindent
The other important difference between a theoretical and an experimental image may be the blurriness of the feature boundaries. The theoretical images always exhibit sharp and well-defined features, whereas an experimental image may consists of features which are slightly spread across their boundaries. In the traditional image processing techniques, blurriness is defined as the case where an image is not well-focused, and each pixel instead of being represented by a single sharp value is ‘mixed’ with neighboring pixels. Such blurring of a sharp image may be simulated via a process of Gaussian blurring, in which an image is convoluted with a kernel that represents a 2D Gaussian distribution with the peak at the center: 

\begin{equation}
	\label{eq:Nonstatic_donor_potential}
	K_{ij} = \frac{1}{N} exp(-\frac{(x^2 + y^2)}{2\sigma^2_B})
\end{equation}

\noindent
Where $x$ and $y$ represents the horizontal and vertical displacement of an element from the center of the kernel respectively, $\sigma_B$ represents the level of blurriness in terms of pixels, and N is the normalization factor to ensure the sum of all elements in the kernel add up to one, such that the resulting blurred image is neither brighter nor darker compared to the original image. The evaluation of the Gaussian blurring kernel is illustrated in the Supplementary Figure~\ref{fig:FigS15}. As an example, Supplementary Figure~\ref{fig:FigS16} shows the kernel values along with the grayscale plot for a 9$\times$9 kernel with $\sigma_B$=2. 
\\ \\
\noindent
The blurring can be made directional and randomized by first making the blurring kernel elliptical, and then rotate it at a random angle:

\begin{equation}
	\label{eq:Nonstatic_donor_potential}
	K_{ij} = \frac{1}{N} exp(-\frac{(xcos\theta + ysin\theta)^2}{\sigma^2_1} - \frac{(ycos\theta - xsin\theta)^2}{\sigma^2_2})
\end{equation} 
\noindent
Where $\sigma_1$ and $\sigma_2$ are independently drawn from the Gaussian distribution of N(0, $\sigma_B$), which when squared results in a $\chi^2$ distribution with 1 degree of freedom, and $\theta$ is drawn from a uniform distribution from [0, 2$\pi$] which represents rotation of the ellipse. The kernels of a randomized Gaussian blurring with $\sigma_B$=3 are shown in the Supplementary Figure~\ref{fig:FigS17}. The Supplementary Figure~\ref{fig:FigS18} plots the STM images corresponding to the ($4,3/4,7,0$) position after the application of various levels of blurring noise. Based on these plots, we estimate that a blurriness noise represented by $\sigma_B \leq$ 3.0 is reasonable to identify STM images with high fidelity and is also commensurate with the level of changes typically observed in the published measurements. 
\\ \\
\noindent
We should also mention here that in the preparation of the training data set, we only considered planar noise. However, the test images included both planner and blurring noise with randomly selected strengths.  

\section*{S3. Edge detection reduction scheme}

\noindent
Supplementary Figure~\ref{fig:FigS19} shows the complete procedure applied for the STM image reduction in case of the edge detection scheme. As an example, an STM image corresponding to the ($4,1/2,8,5$) position is selected and converted to grayscale as shown in part (a). It can be noted that the bright features are present around the center region and a large portion of the image is dark, indicating negligible tunnelling current. The overall size of the image can be reduced by first rotating the image clockwise by 45$^o$ as shown in (b) and then cropping the dark pixels in the image as exhibited in (c). The cropping is done by setting a threshold value for the tunnelling current, below which the image pixels are removed. After these steps, the size of the image is reduced from 535$\times$535 pixels to 237$\times$189 pixels.   
\\ \\
\noindent
We further note that the information is stored in the size and shape of the bright features, therefore the extraction of feature edges will preserve the information about the underpinning donor positions. This step is performed by a mathematical convolution function with the following kernel~\cite{SM}{Ziou98edgedetection}:   

$$
K_{edge} = 
\begin{pmatrix}
-1&-1&-1\\
-1&8&-1\\
-1&-1&-1\\
\end{pmatrix}
$$

\noindent
The convoluted image is plotted in (d). Finally, the size of image can be further reduced by applying a sub-sampling or max-pooling function. This function divides the pixel map into smaller sections and for each section, the maximum value is selected. The image after the application of a 3$\times$3 max-pooling function is shown in (e) which consists of 79$\times$63 pixels. The overall reduction of the STM image size from 535$\times$535 pixels to 79$\times$63 pixels significantly improves the computational efficiency and storage requirements of the neural network in both training and testing phases. We should note here that the final size of the STM image (79$\times$63 pixels) is specific for this particular image selected as an example. Since each STM image consists of different size and number of features, the eventual size of the processed STM will vary from one image to the other image. Based on all the processed images, we found that the final image size is always below 90$\times$90 pixels. 

\section*{S4. Feature averaging reduction scheme}

\noindent
The STM image reduction procedure for the feature averaging scheme is shown in the Supplementary Figure~\ref{fig:FigS20} for an exemplary image corresponding to the position ($4,0,1,8$). The first two steps are same as previously described in the section S3 and are shown in the parts (a) and (b). 
\\ \\
\noindent
After the rotation and cropping of the STM image, we apply the feature averaging procedure. The idea behind this scheme is based on the fact that the image features, which contain the information about the underpinning donor positions are present on atomic positions at the dimer sites. As each feature is distinct in an image and the arrangement of the features is also unique between different donor positions, it is possible to significantly reduce the size of an STM image by only persevering average brightness of each feature based on dimer row atoms. This is done by first overlaying the image with dimer atom positions as shown in part (c). There are 11$\times$10 atom positions which roughly cover all of the visible bright features. Based on the inspection of the computed STM images, the bright features are around the dimer rows, and these features can be represented by simply finding the average brightness of each feature around each dimer atom. This procedure reduces the STM image size to only 11$\times$10 pixels. The image after the application of the feature averaging procedure is illustrated in the part (d). We note that in this image reduction scheme, we do not apply max-pooling operation.

\section*{S5. Training and testing of the convolutional neural network (CNN)}

\noindent
The implementation of the convolutional neural network (CNN) was performed by using Keras~\cite{chollet2015keras} with TensorFlow~\cite{tensorflow2015-whitepaper} as the back-end. For the edge detection scheme (described in the section S3 above), the processed STM images including the planar noise were used to train the CNN. The CNN consisted of a convolutional layer with 32, 3$\times$3 kernels along with 2$\times$2 max-pooling, followed by a hidden layer of 256 rectified linear unit (ReLu) activated neurons as they offer better learning rate for the CNN. Training with 30 epochs yielded a nearly perfect learning and an accuracy of 100\% for a test set consisting of STM images without the addition of noise.
\\ \\
\noindent
In the case of the feature averaging scheme (described in the section S4 above), the CNN consisted of a hidden layer of 64 ReLu activated neurons. The training with 20 epochs led to nearly perfect learning and an accuracy of 100\% for a test set consisting of STM images without the addition of noise. We note here that the number of neurons (256 and 64) are optimised by testing various different numbers, and by selecting the least number of neurons that maintain a near perfect learning.
\\ \\
\noindent
After the training of the CNN, in the next phase, we performed its testing for the images with the application of random levels of noise described in the section S3 above. First, we evaluated the CNN fidelities with the introduction of blurring noise only for both the edge detection and feature averaging schemes. In this case, 16 STM images are arbitrarily chosen (shown in Figure~\ref{fig:FigS21}), with 100 samples of each blurring level ranging from $\sigma_B$=0.5 to 5.0 at an increment of 0.5, establishing a test set of total 17600 images. The accuracy is measured based on the proportion of correct predictions out of the 100 samples. For each test image, the CNN returns a set of probability values indicating the probability of that image being in one of the particular image class. The Supplementary Figure~\ref{fig:FigS22} and~\ref{fig:FigS23} show the graphs of fidelities as a function of the applied blurring level for each of the test class. For low values of $\sigma_B$, the CNN identifies STM images with nearly 100\% accuracy for all the plotted classes. When the blurring noise increases, the accuracy of feature detection scheme is clearly higher than the edge detection scheme. We also note that the decrease in identification accuracy is dependent on the STM image class.  
\\ \\
\noindent
To perform a comparison between the performance of the two image reduction methods, we plot the average fidelities as a function of the blurring noise strength along with the sample standard deviation in the Figure 4(a) of the main text. This plot clearly shows that overall the feature averaging method has a better fidelity compared to the edge detection. This is also accompanied with significantly higher computational efficiency. Based on 10$^5$ training images and 17600 test images, we find that the feature averaging method takes about 30 minutes time, compared to about 3 hour's time frame for the edge detection case on an average desktop machine. Overall, we conclude that the feature averaging method is a better choice for the training a CNN capable of spatial metrology of donor-based qubits in silicon.
\\ \\
\noindent
To test the fidelities of the trained CNN for STM images perturbed by both planar and blurring noise, we apply both noise simultaneously to the STM images plotted in the Supplementary Figure~\ref{fig:FigS21}. Each image is then processed according to both edge detection and feature averaging schemes. After the image processing, the final test images are plotted in Figures 4 (b) and (c) of the main text. In each case, the CNN identifies the donor positions with 100\% accuracy and the corresponding values of the confidence level are also provided in the figure. 

\clearpage
\newpage

\renewcommand{\thefigure}{\textbf{S\arabic{figure}}}
\renewcommand{\figurename}{\textbf{Supplementary Fig.}}

\renewcommand{\thetable}{\textbf{T\arabic{table}}}
\renewcommand{\tablename}{\textbf{Supplementary table.}}

\setcounter{figure}{0}

\begin{figure*}
\includegraphics[scale=0.43]{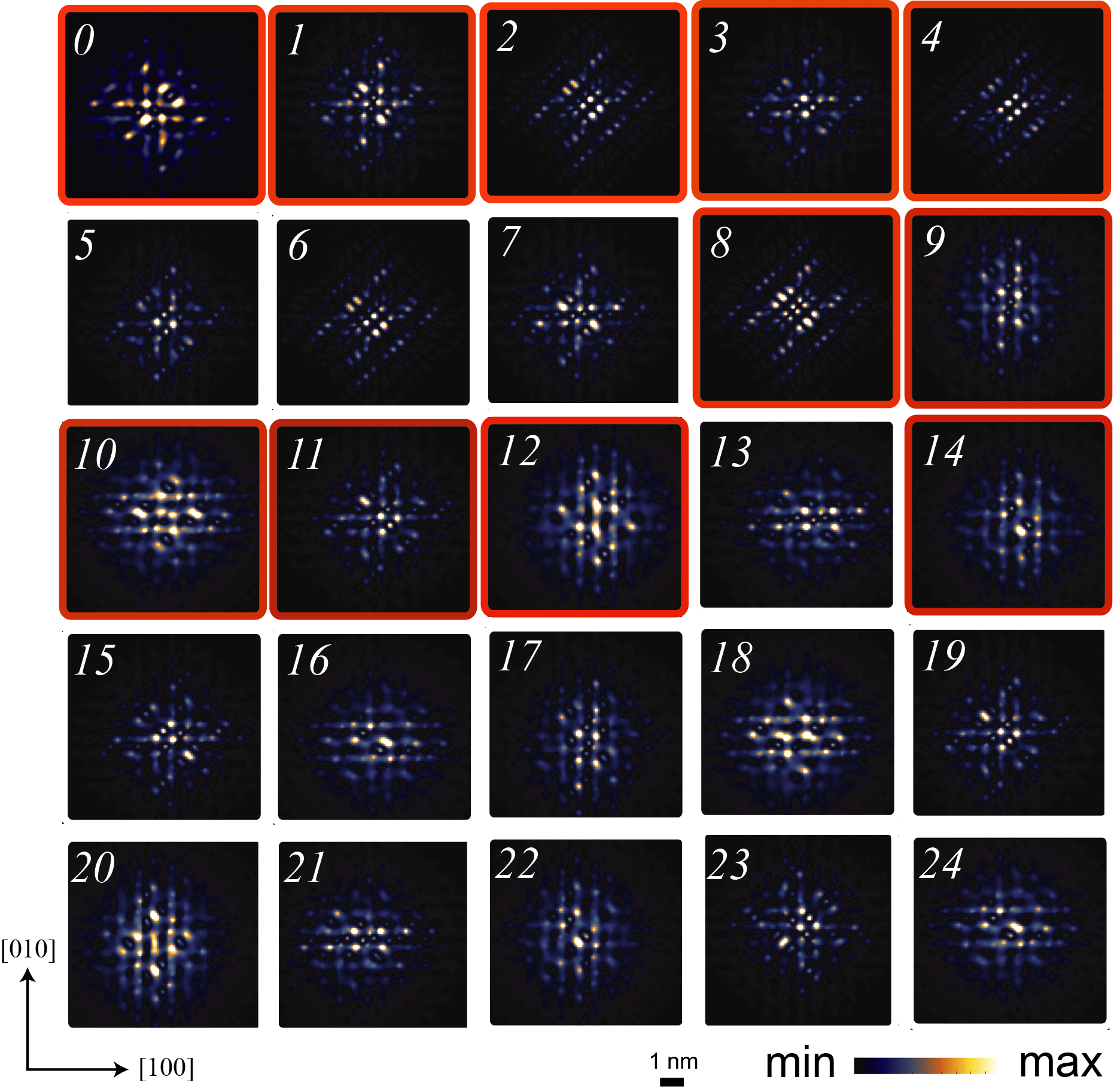}
\caption{Theoretically computed STM images are shown for all possible positions ($j$=0, 1, 2, ..., 24) at $n$=4, $m$=0, and $i$=1. The images clearly exhibit well-defined symmetry of the wave functions convoluted with surface dimer positions. Based on symmetry analysis, we find that the 2P images are identical when the second phosphorus atom is symmetrically distributed around the reference phosphorus atom at $j$=0. The distinct images are highlighted by the red border.}
\label{fig:FigS1}
\end{figure*}

\begin{figure*}
\includegraphics[scale=0.43]{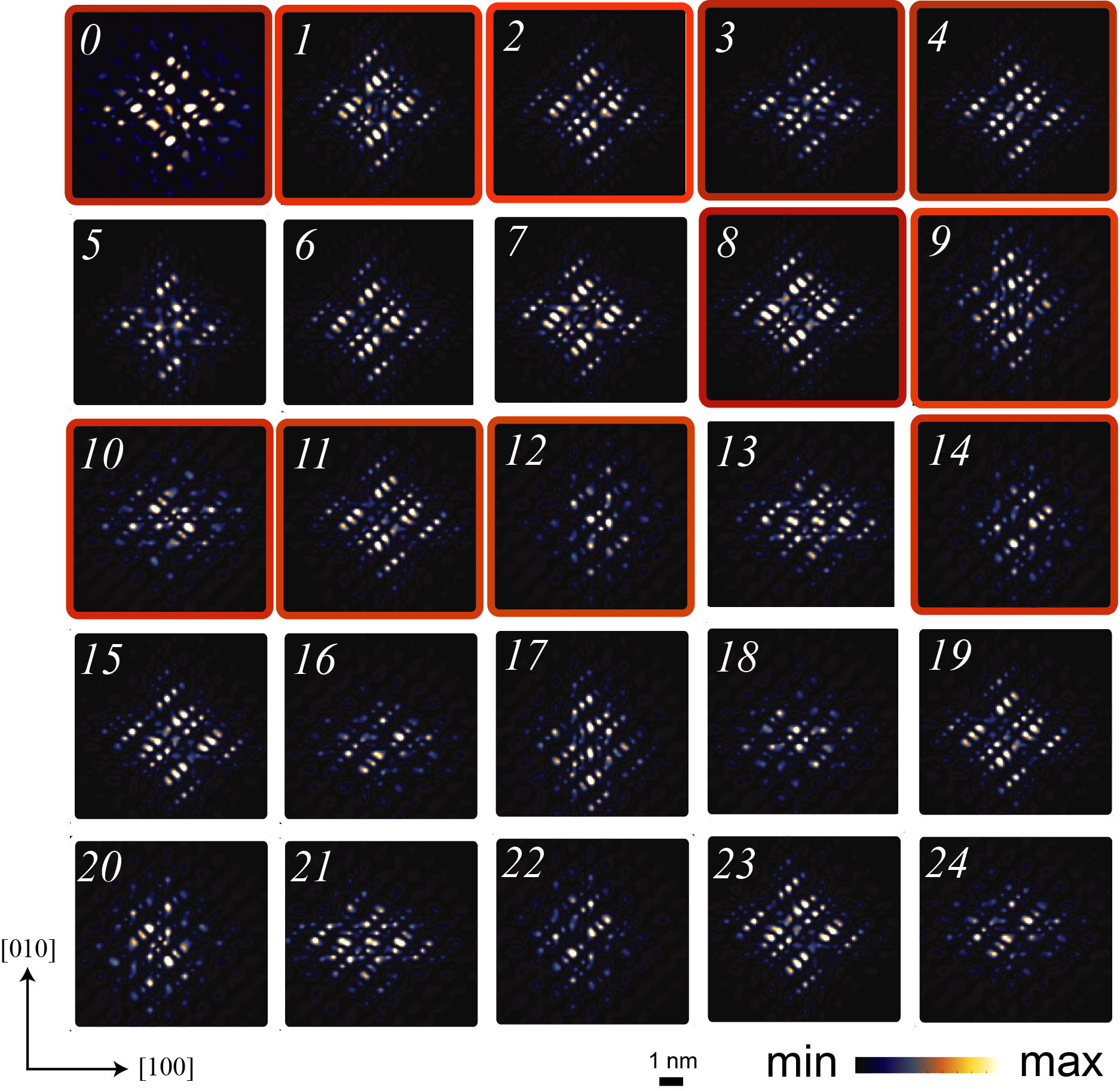}
\caption{Theoretically computed STM images are shown for all possible positions ($j$=0, 1, 2, ..., 24) at $n$=4, $m$=1/4, and $i$=3. The images clearly exhibit well-defined symmetry of the wave functions convoluted with surface dimer positions. Based on symmetry analysis, we find that the 2P images are identical when the second phosphorus atom is symmetrically distributed around the reference phosphorus atom at $j$=0. The distinct images are highlighted by the red border.}
\label{fig:FigS2}
\end{figure*}

\begin{figure*}
\includegraphics[scale=0.43]{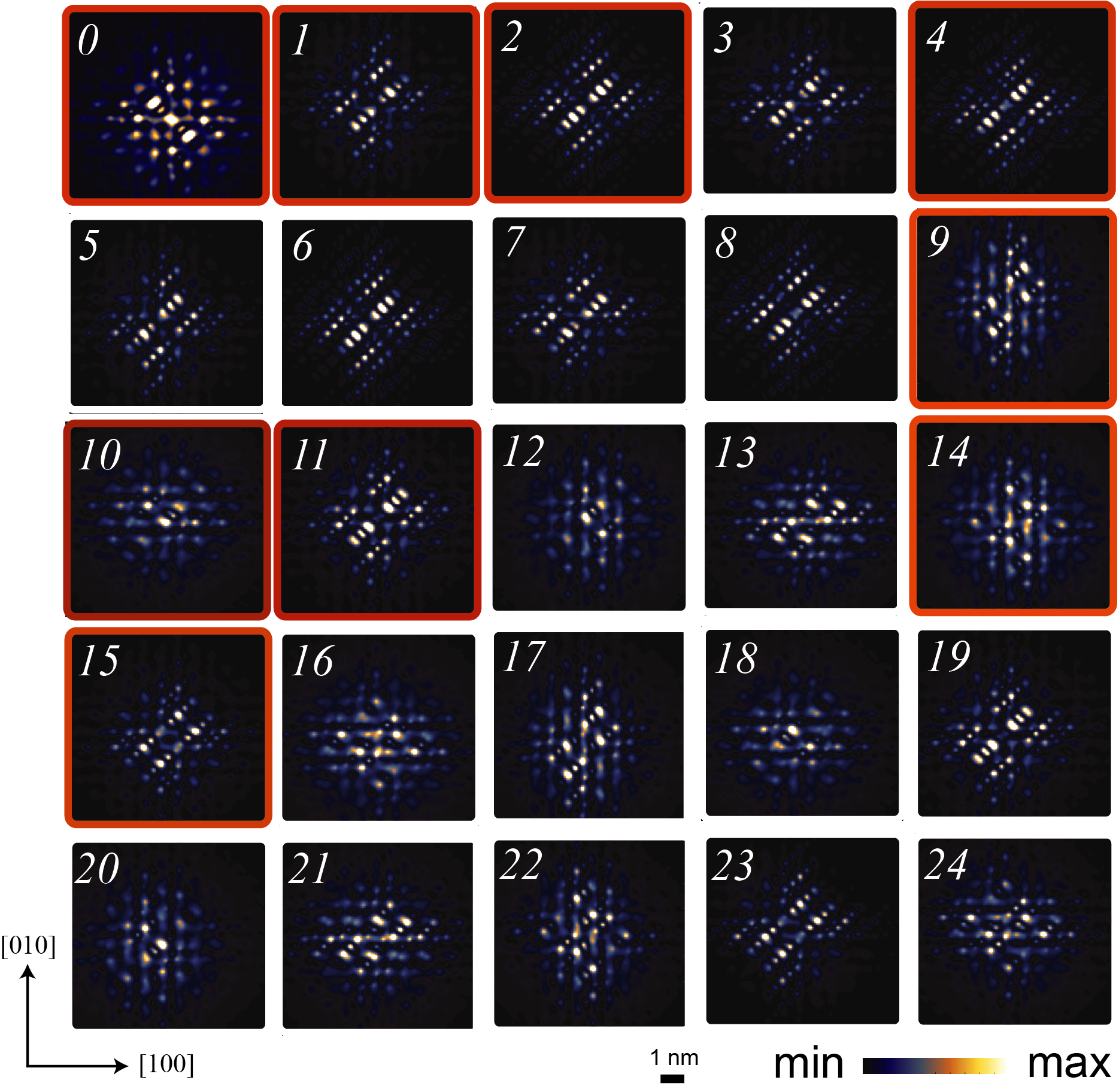}
\caption{Theoretically computed STM images are shown for all possible positions ($j$=0, 1, 2, ..., 24) at $n$=4, $m$=1/2, and $i$=5. The images clearly exhibit well-defined symmetry of the wave functions convoluted with surface dimer positions. Based on symmetry analysis, we find that the 2P images are identical when the second phosphorus atom is symmetrically distributed around the reference phosphorus atom at $j$=0. The distinct images are highlighted by the red border.}
\label{fig:FigS3}
\end{figure*}

\begin{figure*}
\includegraphics[scale=0.43]{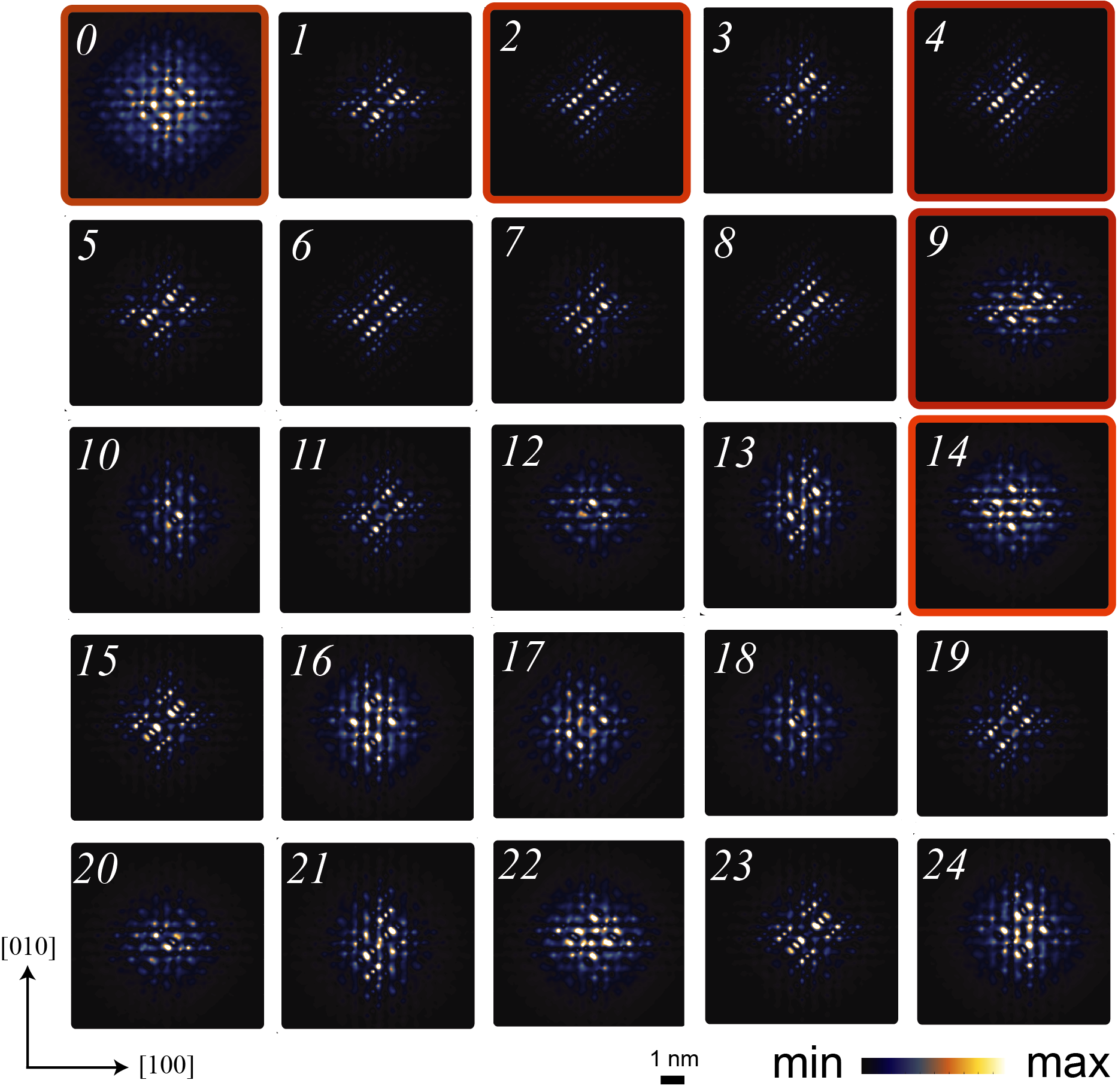}
\caption{Theoretically computed STM images are shown for all possible positions ($j$=0, 1, 2, ..., 24) at $n$=4, $m$=1/4, and $i$=6. The images clearly exhibit well-defined symmetry of the wave functions convoluted with surface dimer positions. Based on symmetry analysis, we find that the 2P images are identical when the second phosphorus atom is symmetrically distributed around the reference phosphorus atom at $j$=0. The distinct images are highlighted by the red border.}
\label{fig:FigS4}
\end{figure*}

\begin{figure*}
\includegraphics[scale=0.43]{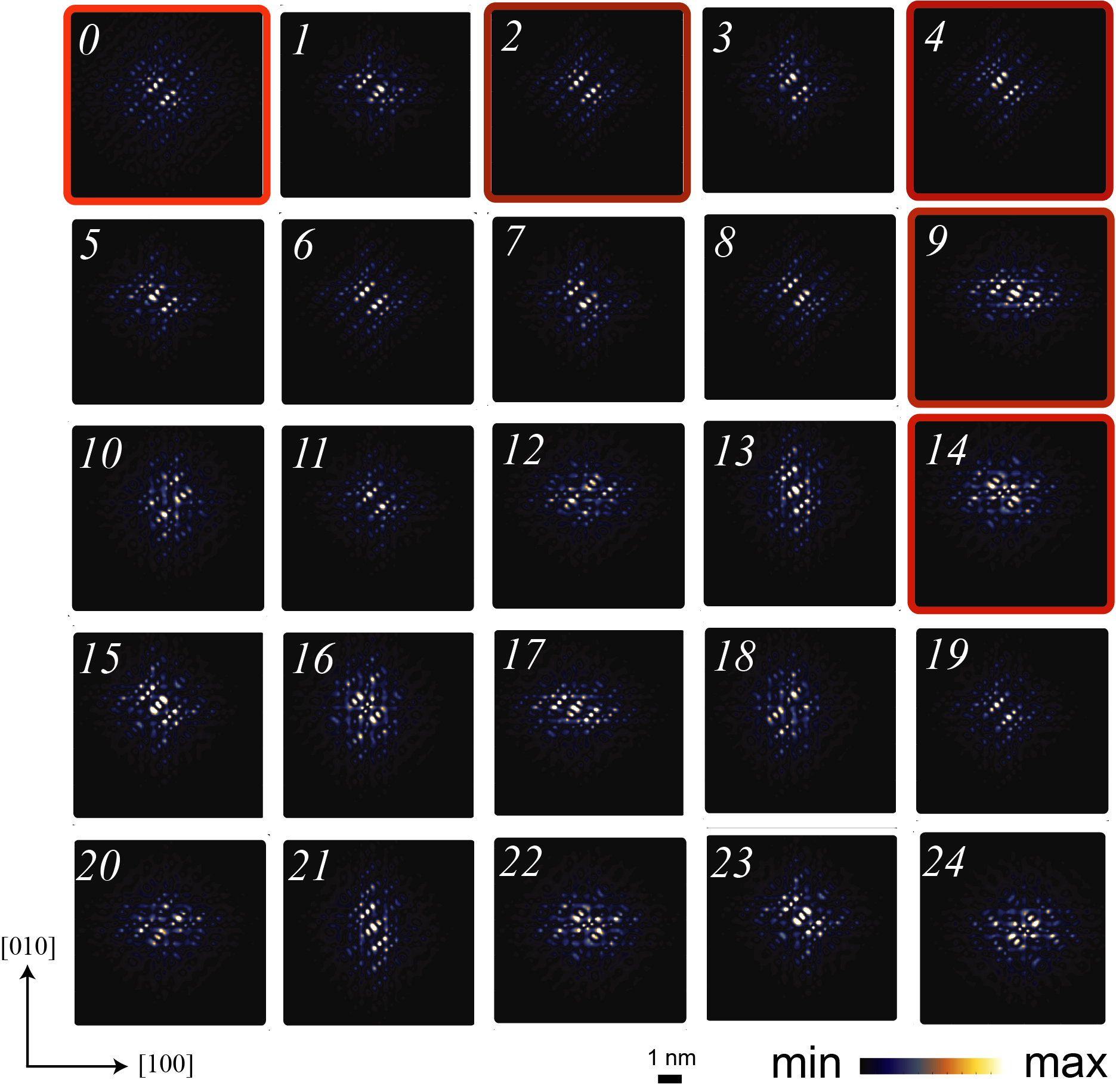}
\caption{Theoretically computed STM images are shown for all possible positions ($j$=0, 1, 2, ..., 24) at $n$=4, $m$=3/4, and $i$=8. The images clearly exhibit well-defined symmetry of the wave functions convoluted with surface dimer positions. Based on symmetry analysis, we find that the 2P images are identical when the second phosphorus atom is symmetrically distributed around the reference phosphorus atom at $j$=0. The distinct images are highlighted by the red border.}
\label{fig:FigS5}
\end{figure*}

\clearpage

\begin{figure*}
\includegraphics[scale=0.43]{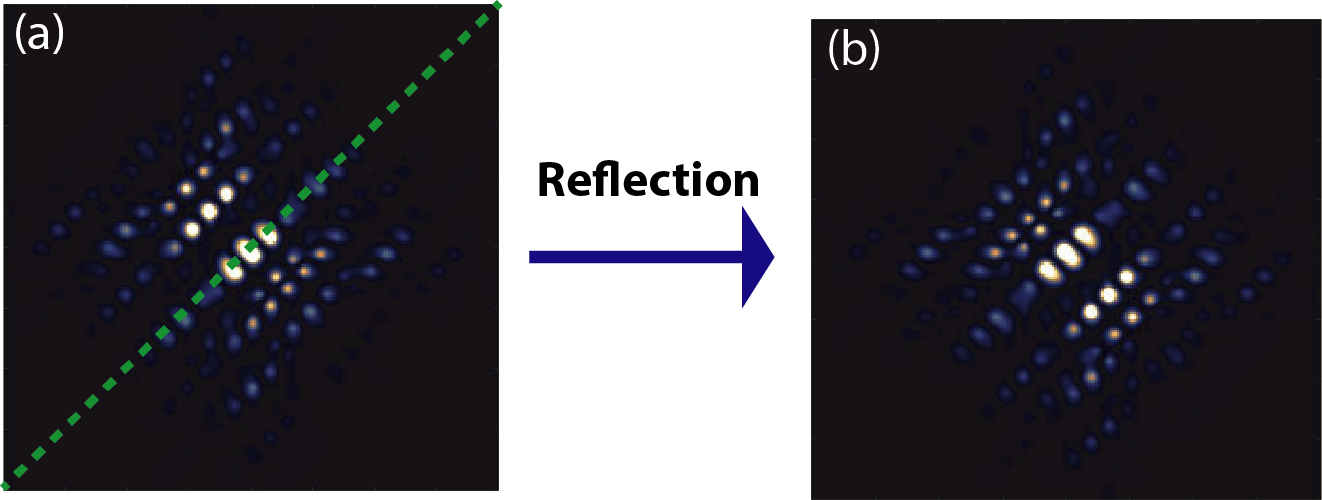}
\caption{An example case is shown in (a) for an STM image corresponding to $n$=4, $m$=3/4, $i$=7, and $j$=4. The green dotted line indicates the diagonal axis parallel to the surface dimer rows. The image is reflected about the green axis in (b), which is now same as the STM image computed for the position corresponding to $n$=4, $m$=3/4, $i$=7, and $j$=8.}
\label{fig:FigS6}
\end{figure*}

\begin{table}[ht]
\centering
\caption{STM images for ($4,3/4,7,j$) positions are sorted in classes of unique images, where each class is defined by ($m,n,i,min(j)$) where $min(j)$ is the minimum value of $j$ in that class. There are in total nine classes.} \label{tableS1}
\includegraphics[scale=0.5]{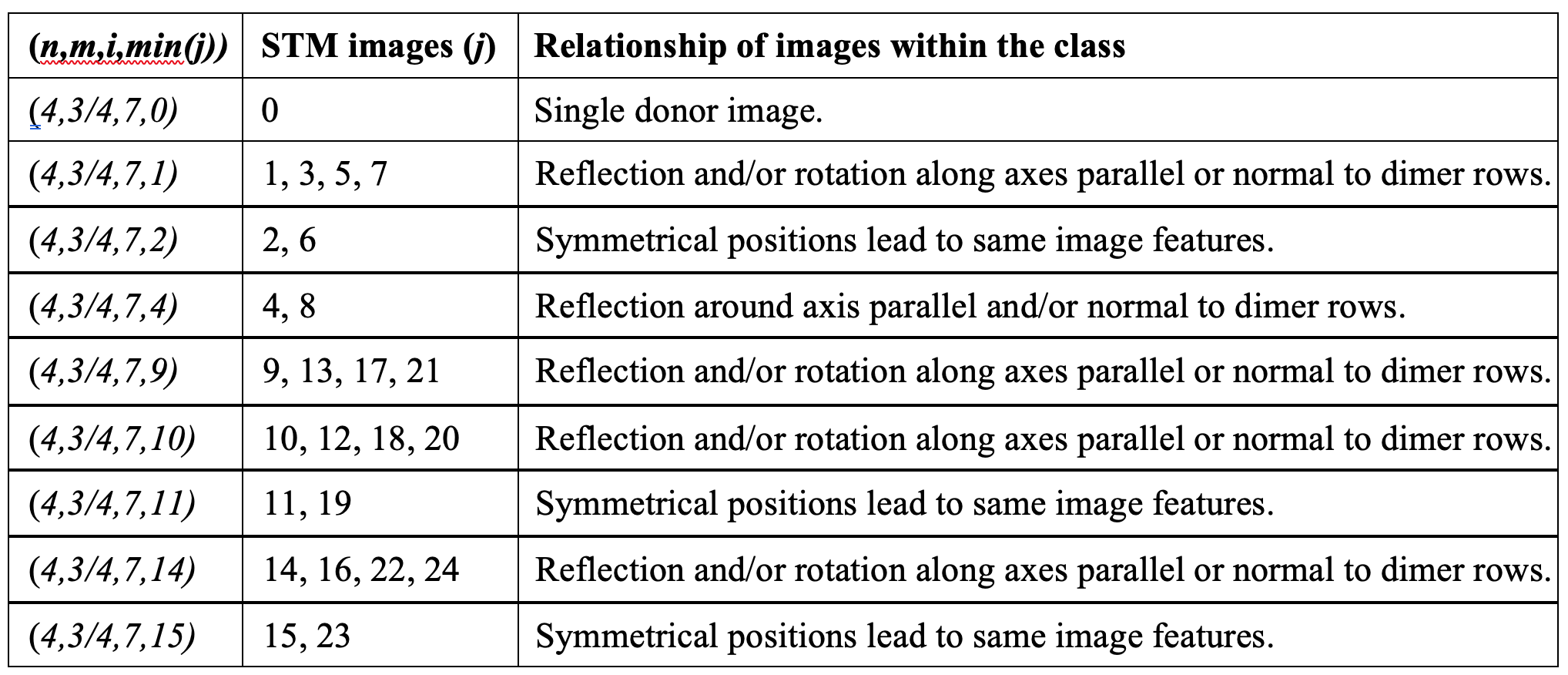}
\end{table}

\begin{table}[ht]
\centering
\caption{STM images for ($4,1/2,5,j$) positions are sorted in classes of unique images, where each class is defined by ($m,n,i,min(j)$) where $min(j)$ is the minimum value of $j$ in that class. There are in total nine classes.} \label{tableS2}
\includegraphics[scale=0.5]{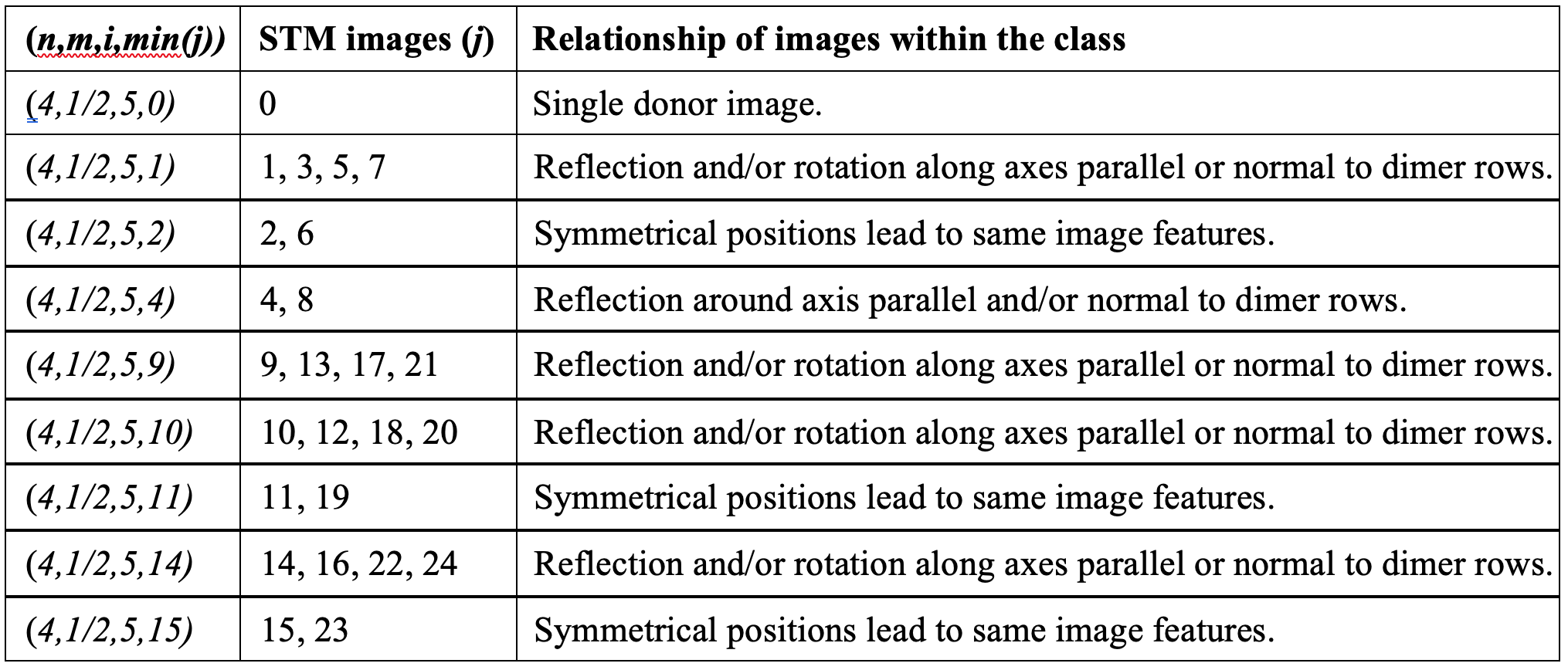}
\end{table}

\begin{table}[ht]
\centering
\caption{STM images for ($4,0,1,j$) positions are sorted in classes of unique images, where each class is defined by ($m,n,i,min(j)$) where $min(j)$ is the minimum value of $j$ in that class. There are in total eleven classes.} \label{tables3}
\includegraphics[scale=0.5]{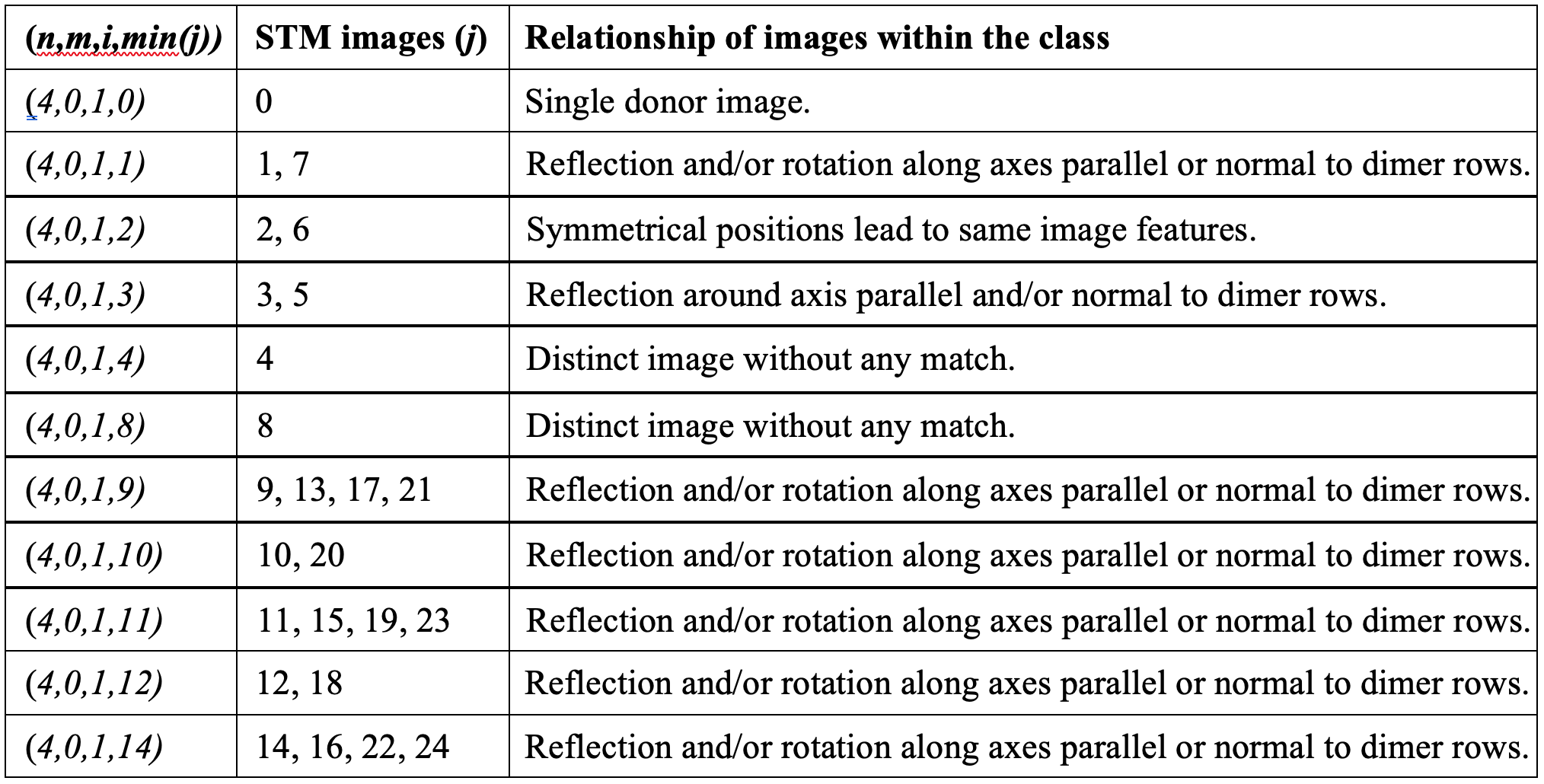}
\end{table}

%\clearpage

\begin{table}[ht]
\centering
\caption{STM images for ($4,1/4,3,j$) positions are sorted in classes of unique images, where each class is defined by ($m,n,i,min(j)$) where $min(j)$ is the minimum value of $j$ in that class. There are in total eleven classes.} \label{tables4}
\includegraphics[scale=0.5]{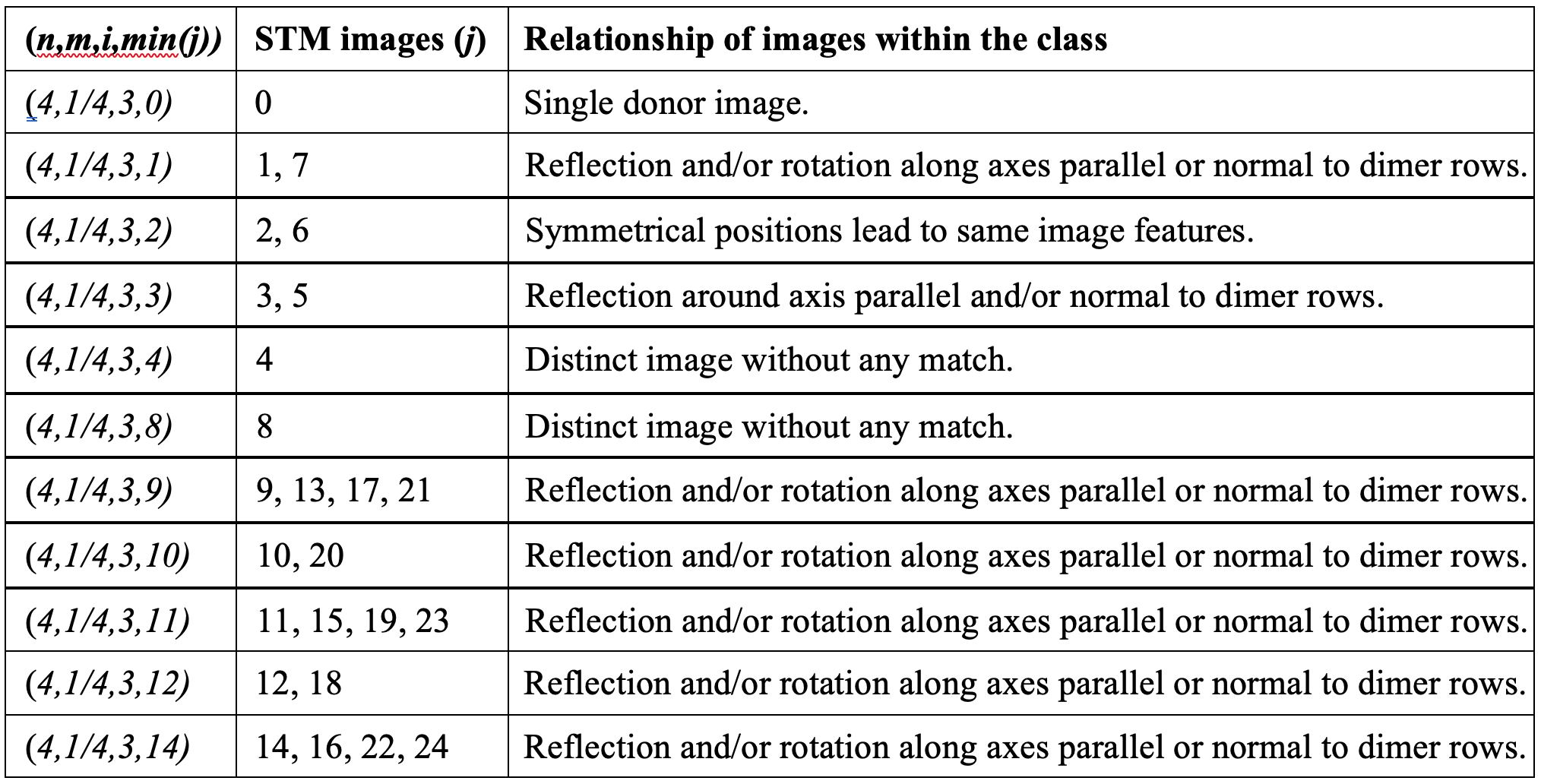}
\end{table}

\begin{table}[ht]
\centering
\caption{STM images for ($4,1/2,6,j$) positions are sorted in classes of unique images, where each class is defined by ($m,n,i,min(j)$) and $min(j)$ is the minimum value of $j$ in that class. Note that in this case, some of the classes consists images which are same as in classes ($4,1/2,5,j$) and therefore are not considered unique. There are in total five classes of unique images.} \label{tables5}
\includegraphics[scale=0.5]{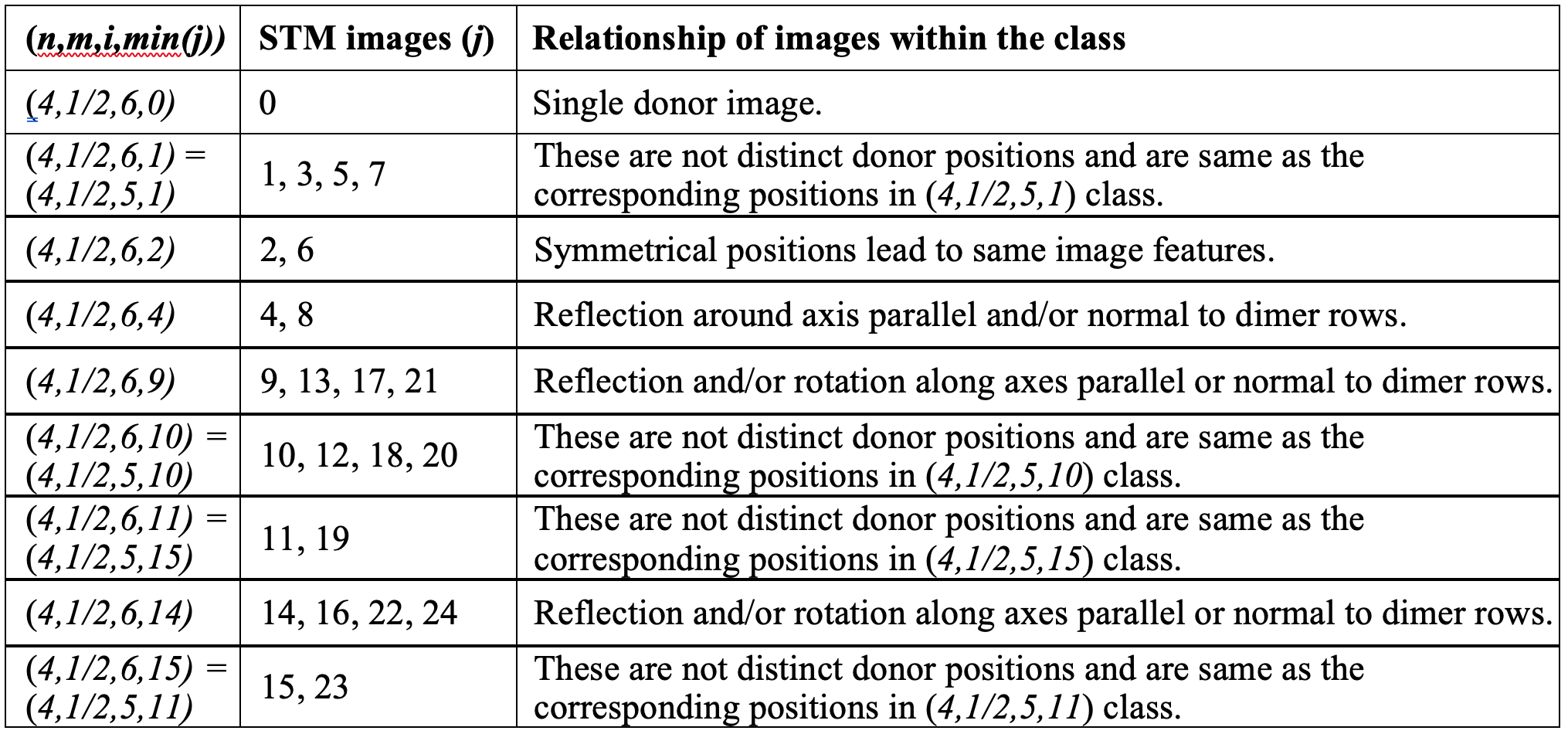}
\end{table}

\begin{table}[ht]
\centering
\caption{STM images for ($4,3/4,8,j$) positions are sorted in classes of unique images, where each class is defined by ($m,n,i,min(j)$) and $min(j)$ is the minimum value of $j$ in that class. Note that in this case, some of the classes consists images which are same as in classes ($4,3/4,7,j$) and therefore are not considered unique. There are in total five classes of unique images.} \label{tables6}
\includegraphics[scale=0.5]{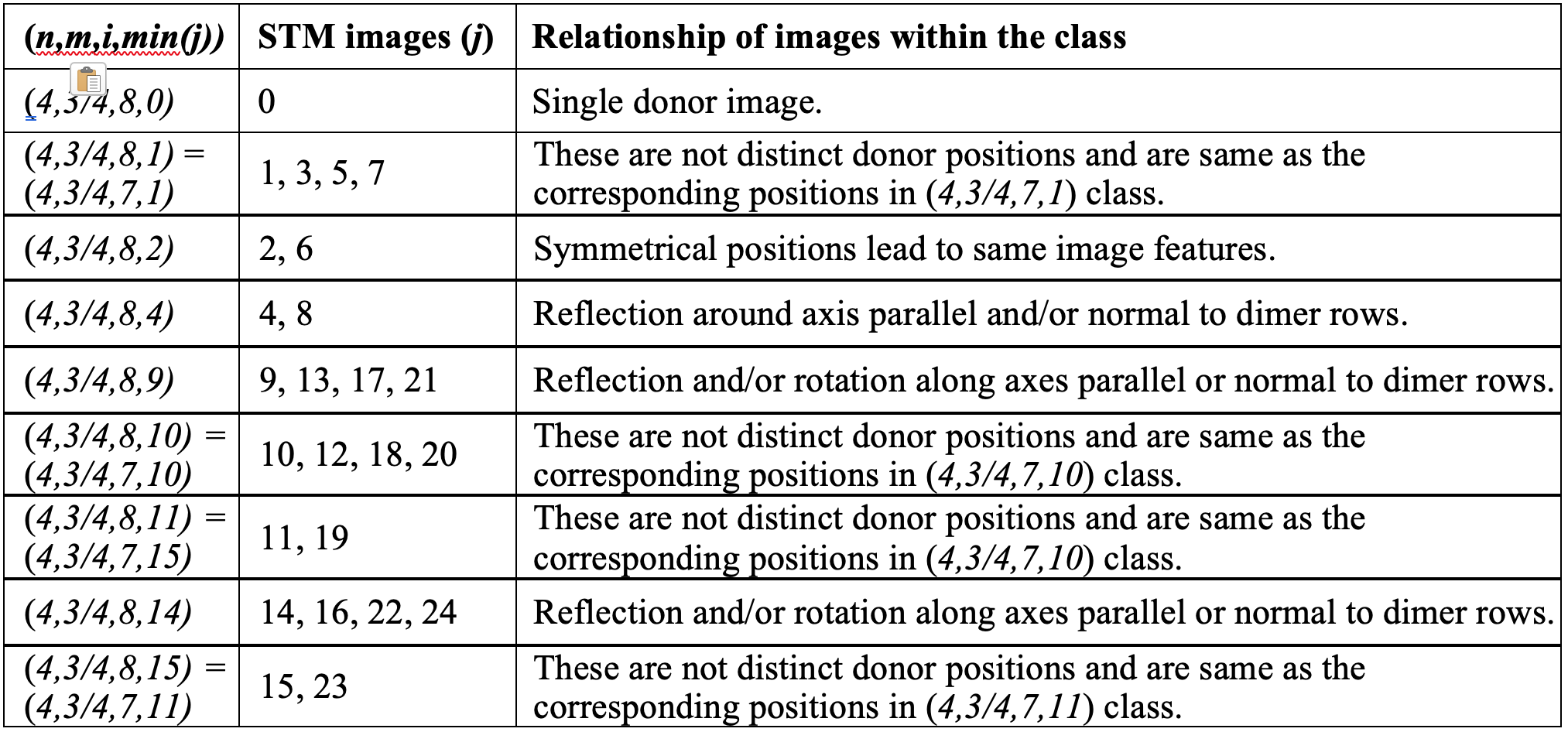}
\end{table}

\begin{figure*}
\includegraphics[scale=1]{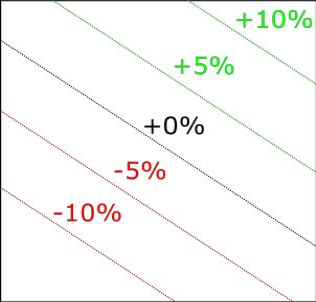}
\caption{Example of planar variation, where contours of percentage variation are plotted. On the upper side the image is brighter, while the lower side the image is darker.}
\label{fig:FigS13}
\end{figure*}

\begin{figure*}
\includegraphics[scale=0.6]{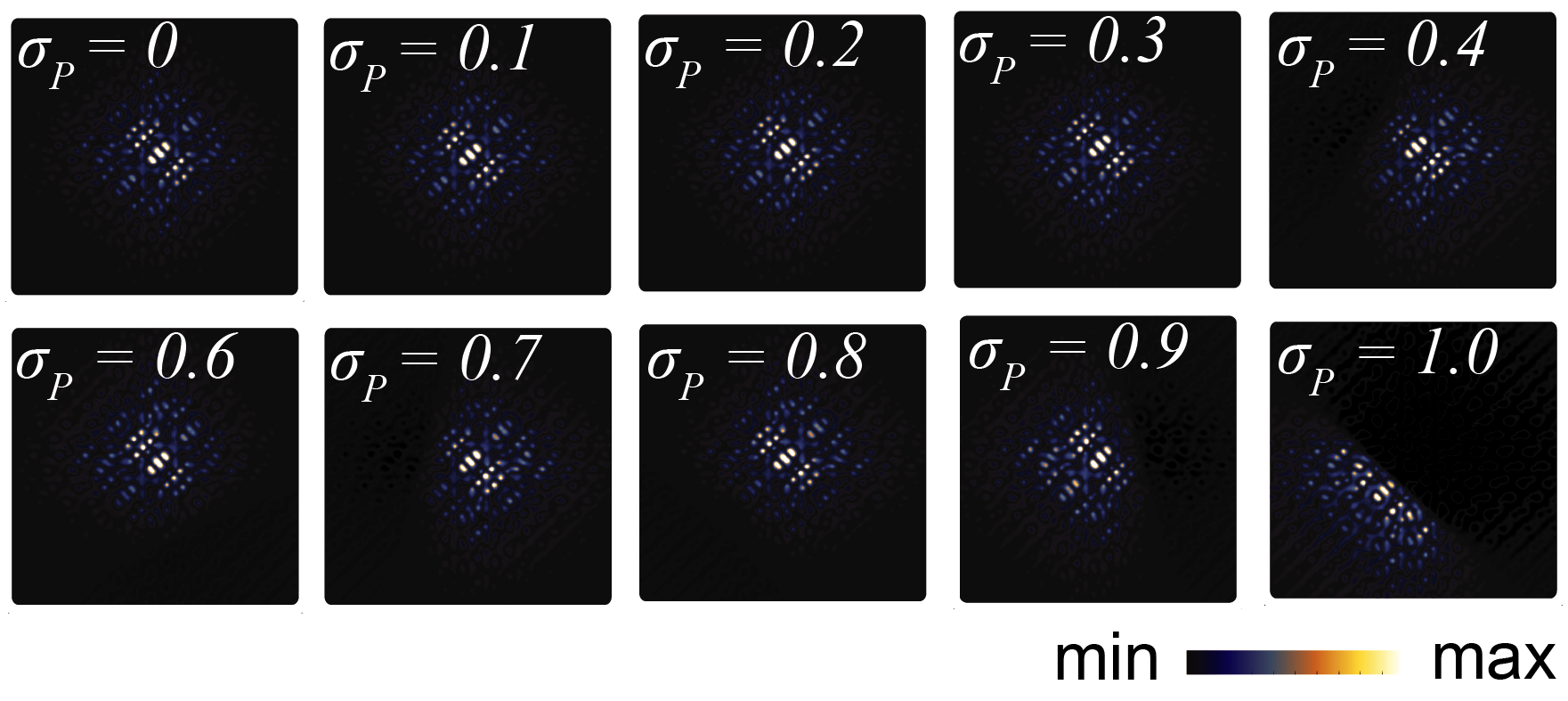}
\caption{An exemplary STM image corresponding to $n$=4, $m$=3/4, $i$=7, and $j$=0 is shown for various planar noise levels ($\sigma_P$). The axis of the planar noise is randomly selected in each case.}
\label{fig:FigS14}
\end{figure*}

\begin{figure*}
\includegraphics[scale=0.45]{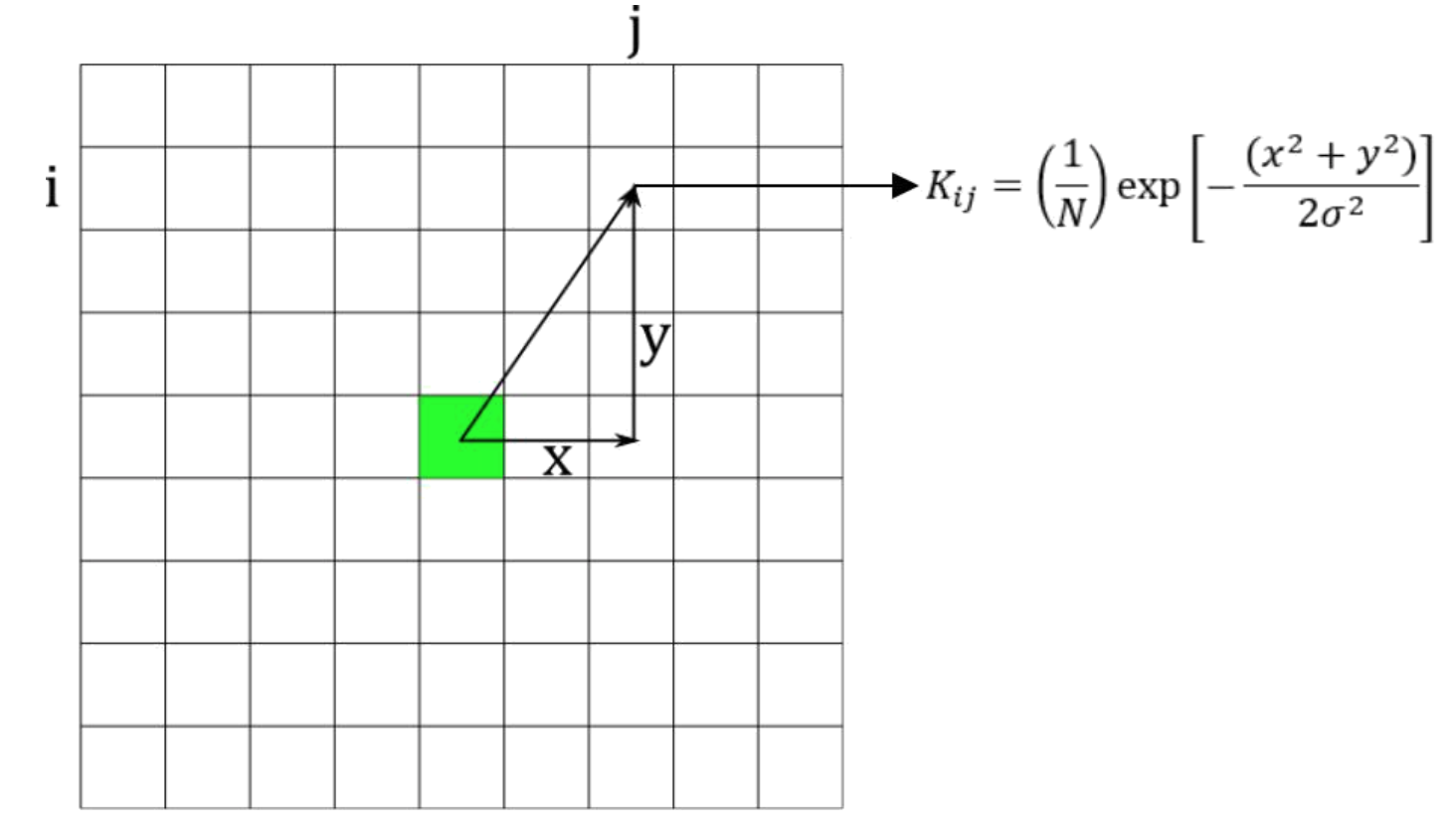}
\caption{The evaluation of the Gaussian blurring kernel, where $x$ and $y$ are the displacements from the center, with $\sigma_B$ is the blurring level based on the number of pixels.}
\label{fig:FigS15}
\end{figure*}

\begin{figure*}
\includegraphics[scale=0.43]{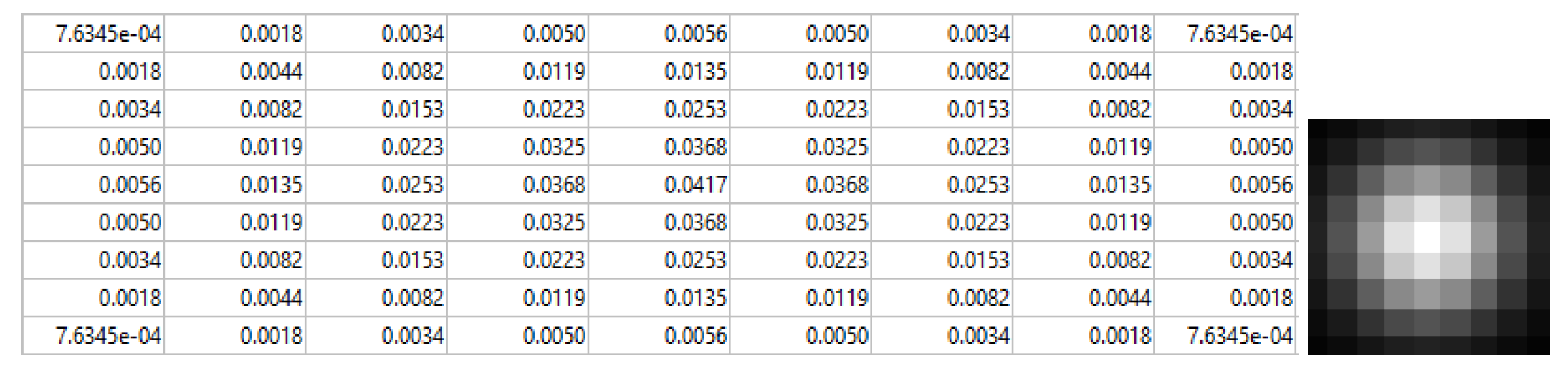}
\caption{An example of a 9$\times$9 Gaussian blurring kernel with $\sigma_B$=2 along with its image plot on the right side.}
\label{fig:FigS16}
\end{figure*}

\begin{figure*}
\includegraphics[scale=0.43]{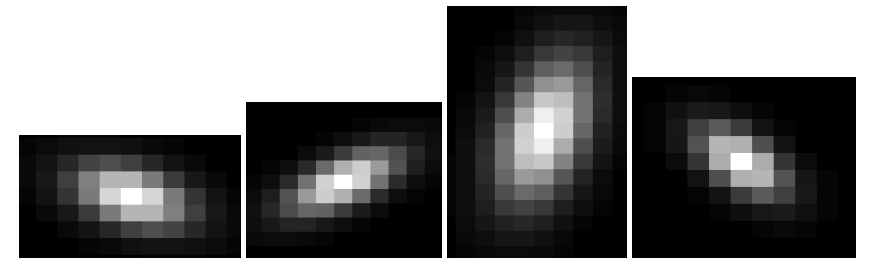}
\caption{Four different samples for the randomized Gaussian blurring kernels for $\sigma_B$=3.}
\label{fig:FigS17}
\end{figure*}

\begin{figure*}
\includegraphics[scale=0.6]{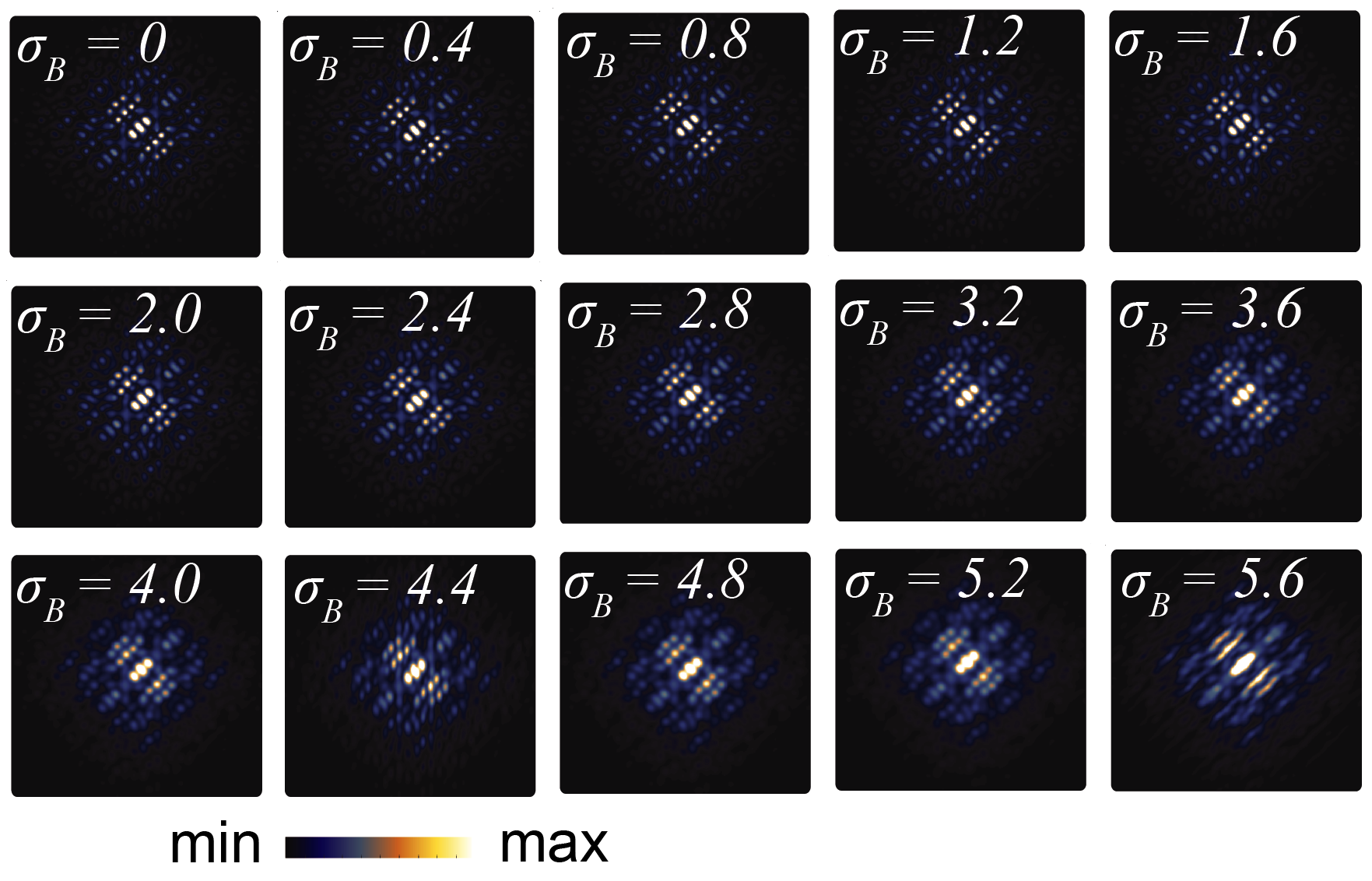}
\caption{An exemplary STM image is shown for position ($4,3/4,7,0$) with various levels of blurring noise. We estimate that a typical measured STM image exhibits a blurring noise varying from 0 to 4.4.}
\label{fig:FigS18}
\end{figure*}

\begin{figure*}
\includegraphics[scale=0.43]{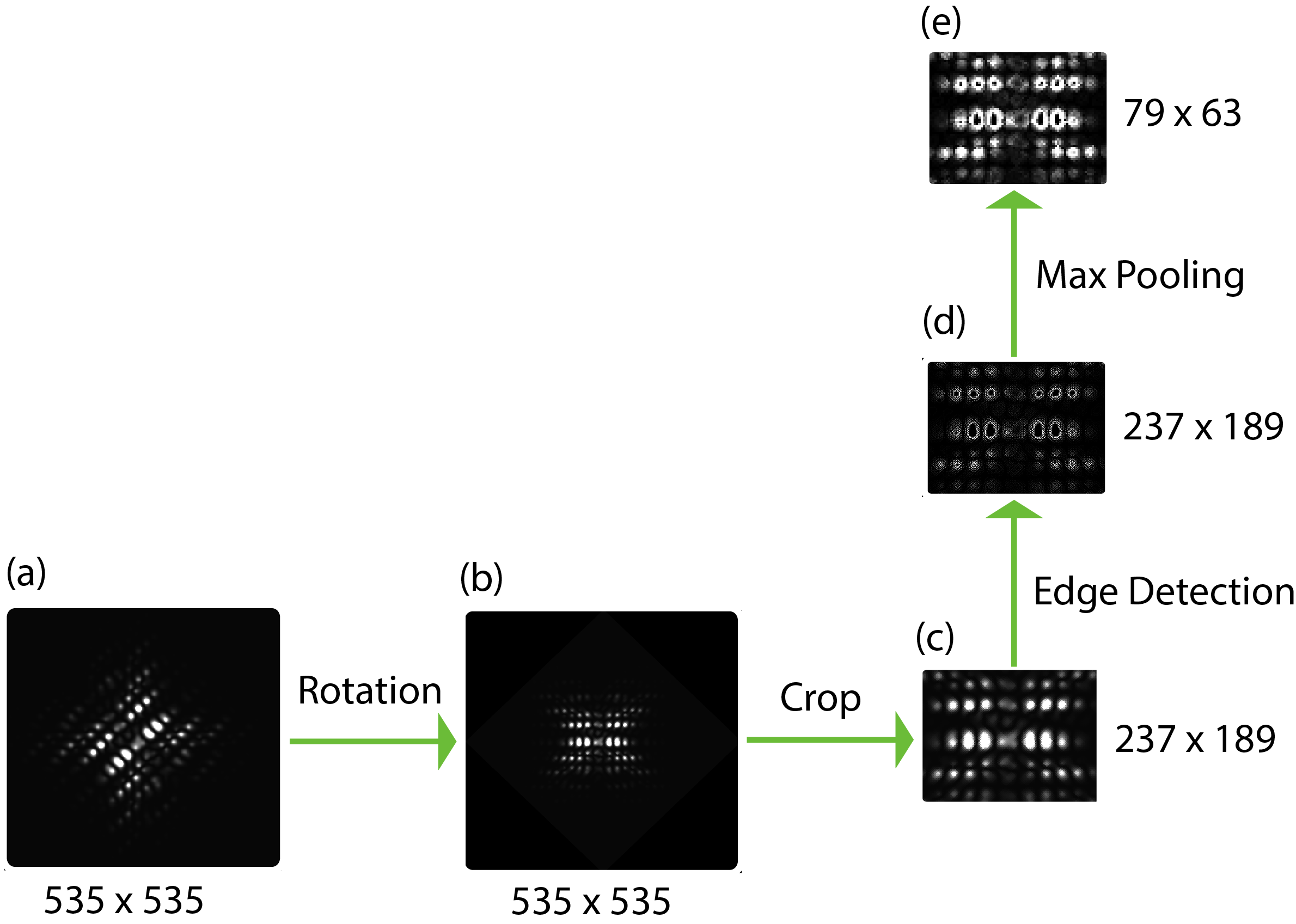}
\caption{Flow chart diagram of STM image processing for the edge detection case is shown for a sample image corresponding to the position ($4,1/2,8,5$). The original gray scale image is 535$\times$535 pixel size. The image is rotated clockwise by 45$^o$ and the black space region (negligible tunneling current) is cropped. The new image is 237$\times$189 pixels size and only consists of bright features in the image. The application edge detection operation is applied to highlight the edges of the bright features. In the final step, a max-pooling function is applied to further reduce the size of image to 79$\times$63 pixels.}
\label{fig:FigS19}
\end{figure*}

\begin{figure*}
\includegraphics[scale=0.65]{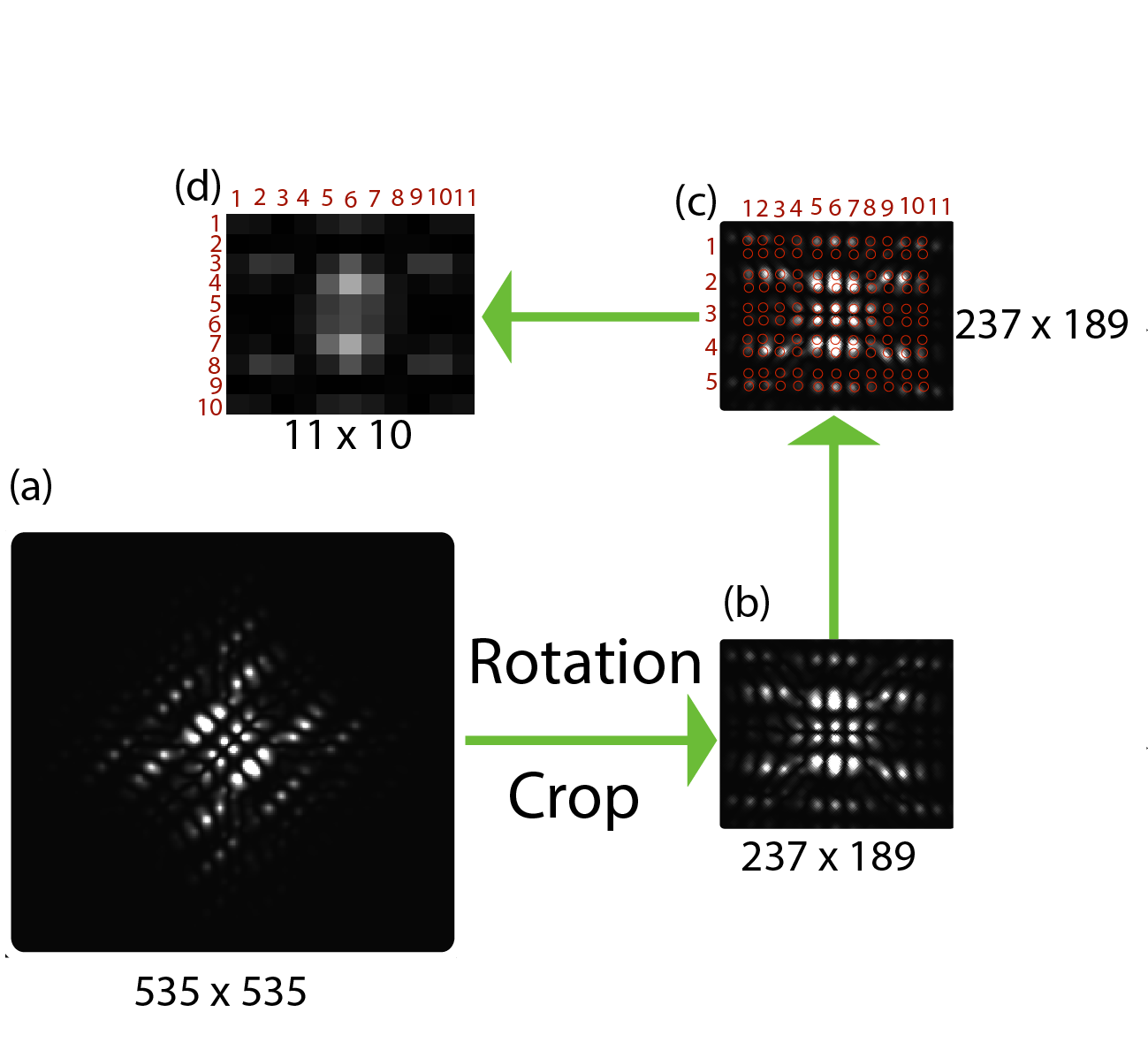}
\caption{Flow chart diagram of feature averaging procedure is shown for a sample STM image corresponding to position ($4,0,1,8$). A grayscale STM image consists of 535$\times$535 pixels. The image is rotated clockwise by 45$^o$ and the black pixels corresponding to zero tunneling current are cropped, reducing the size of STM image to 237$\times$189. The bright features correspond to the position of surface silicon atom in the form of dimer rows. The red circles indicate the positioning of dimer rows with respect to the bright features in the STM image. For this particular STM image, there are 11$\times$10 dimer row atoms shown by the red circles. The average value of pixels in each red circle is computed to form a new image representation consisting of only 11$\times$10 pixels.}
\label{fig:FigS20}
\end{figure*}

\clearpage

\begin{figure*}
\includegraphics[scale=0.43]{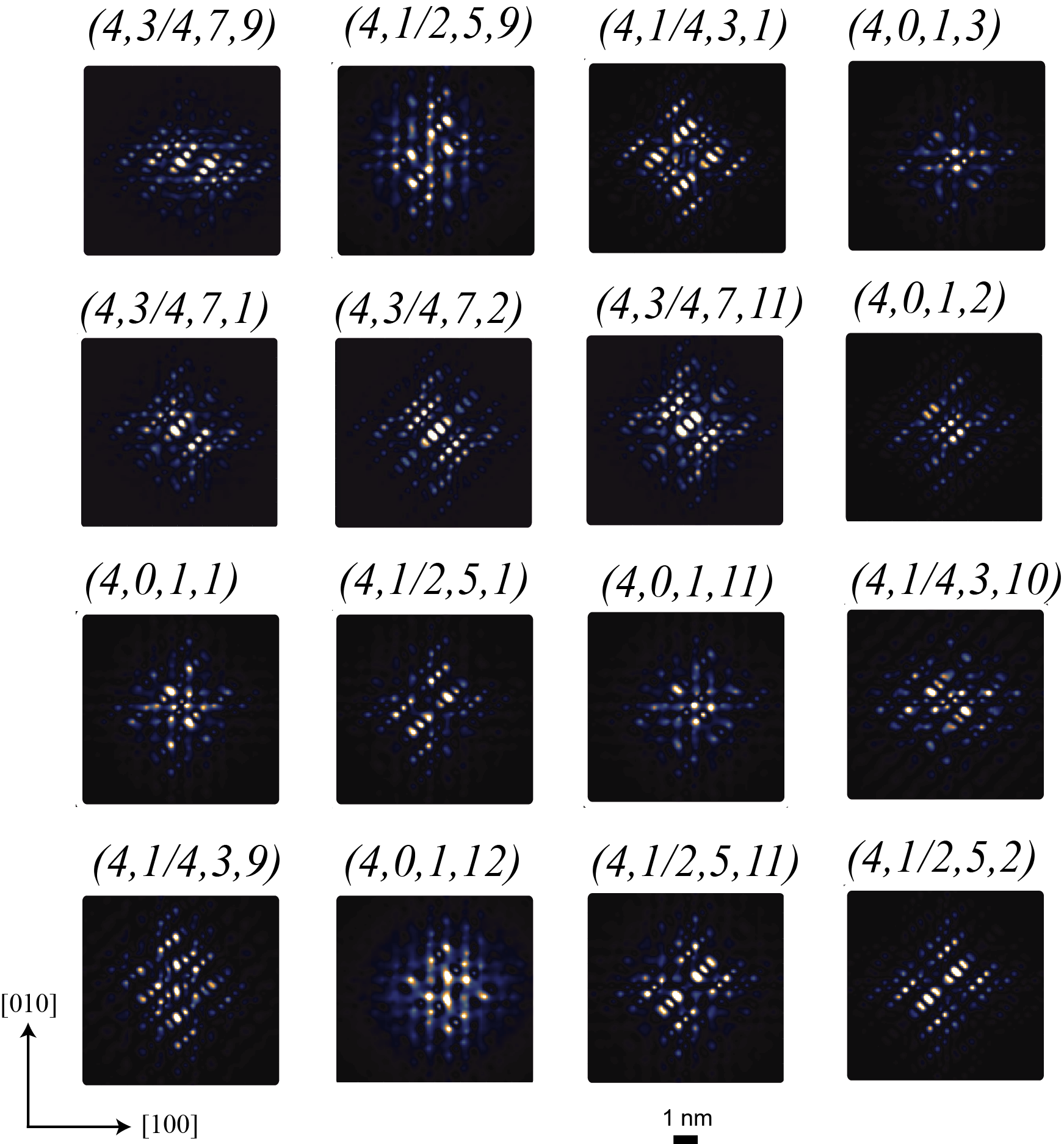}
\caption{A set of sixteen STM images corresponding to donor locations marked with ($n,m,i,j$) to test the fidelity of machine learning framework. Each image is processed in accordance with procedures described in sections S3 and S4 and perturbed with random planar and blurring noise values according to the description in section S2. The final processed images are shown in Figure 4 (b) and (c) of the main manuscript.}
\label{fig:FigS21}
\end{figure*}

\begin{figure*}
\includegraphics[scale=0.43]{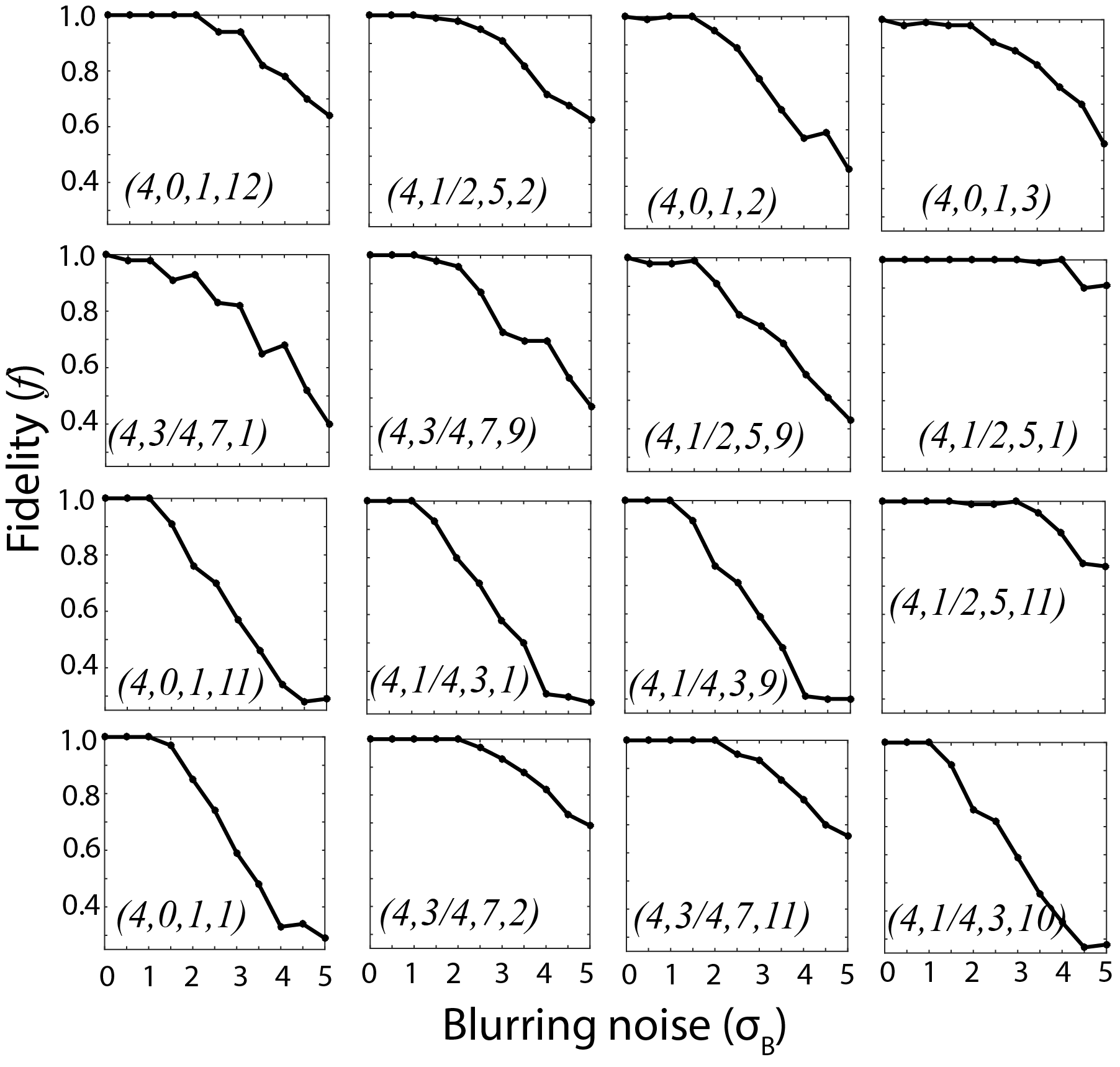}
\caption{Plots of fidelities ratios based on 100 images with random blurring noise as a function of the blurring noise for the set of selected 16 images corresponding to donor positions identified by ($n,m,i,j$). The images are processed based on the edge detection scheme. Our results show that the rate of fidelity drop is different for different STM image classes at the same noise level, indicating that the machine learning identification will be dependent on the target depth in the qubit fabrication process.}
\label{fig:FigS22}
\end{figure*}

\begin{figure*}
\includegraphics[scale=0.43]{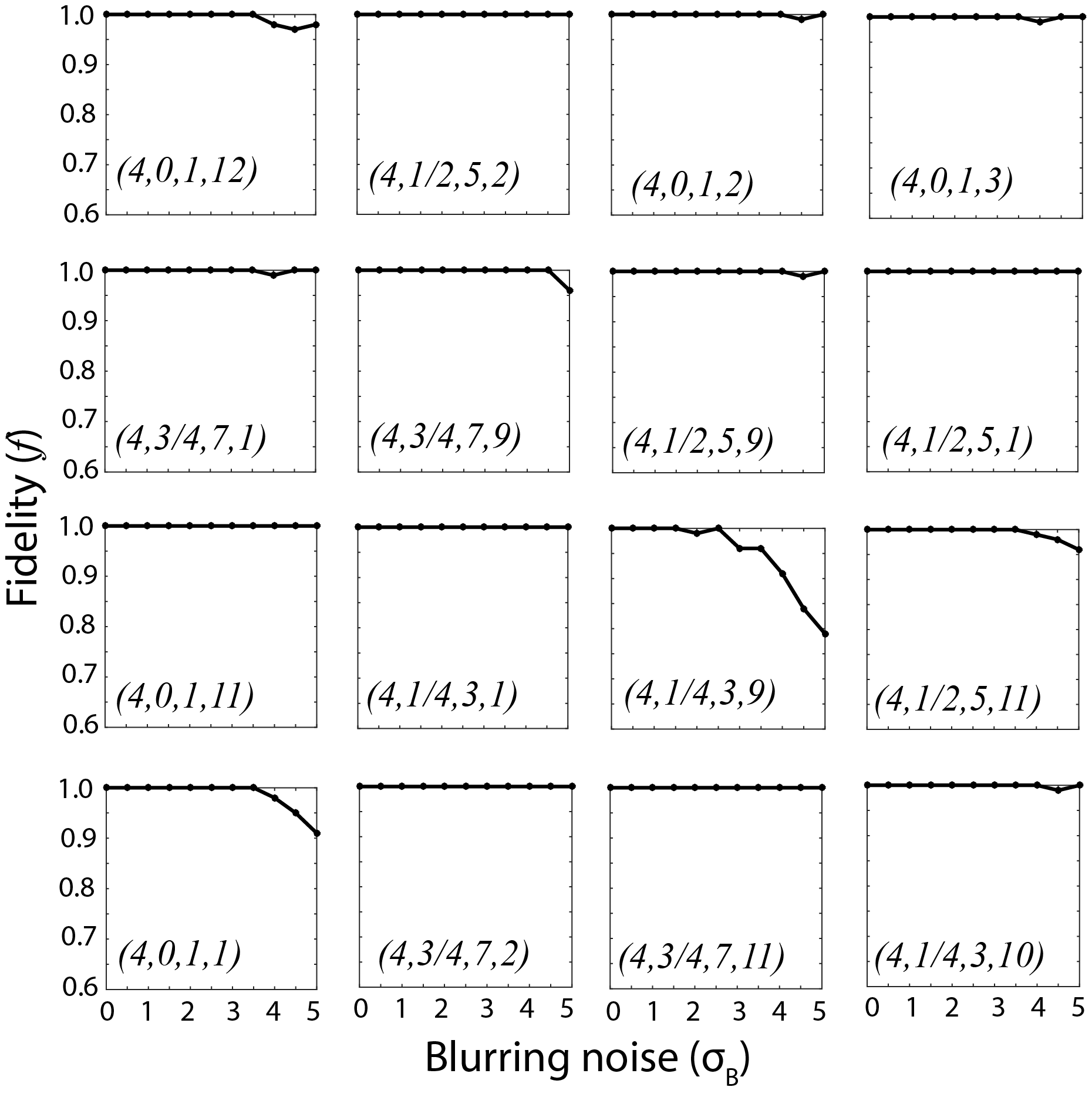}
\caption{Plots of fidelities ratios based on 100 images with random blurring noise as a function of the blurring noise for the set of selected 16 donor images corresponding to donor positions identified by ($n,m,i,j$). The images are processed based on feature average processing scheme.}
\label{fig:FigS23}
\end{figure*}

\clearpage
\newpage

\def\bibsection{\subsection*{\refname}}

\bibliographystyle{naturemag}
%\bibliography{SiDonor}

\begin{thebibliography}{10}
\expandafter\ifx\csname url\endcsname\relax
  \def\url#1{\texttt{#1}}\fi
\expandafter\ifx\csname urlprefix\endcsname\relax\def\urlprefix{URL }\fi
\providecommand{\bibinfo}[2]{#2}
\providecommand{\eprint}[2][]{\url{#2}}

\bibitem{Kane_Nature_1998}
\bibinfo{author}{Kane, B.~E.}
\newblock \bibinfo{title}{A silicon-based nuclear spin quantum computer}.
\newblock \emph{\bibinfo{journal}{Nature}} \textbf{\bibinfo{volume}{393}},
  \bibinfo{pages}{133} (\bibinfo{year}{1998}).

\bibitem{Loss_PRA_1998}
\bibinfo{author}{Loss, D.} \& \bibinfo{author}{Divincenzo, D.}
\newblock \bibinfo{title}{Quantum computation with quantum dots}.
\newblock \emph{\bibinfo{journal}{Phys. Rev. A}} \textbf{\bibinfo{volume}{57}},
  \bibinfo{pages}{120} (\bibinfo{year}{1998}).

\bibitem{Zwanenburg_RMP_2013}
\bibinfo{author}{Zwanenburg, F.} \emph{et~al.}
\newblock \bibinfo{title}{Silicon quantum electronics}.
\newblock \emph{\bibinfo{journal}{Rev. Mod. Phys.}}
  \textbf{\bibinfo{volume}{85}}, \bibinfo{pages}{961} (\bibinfo{year}{2013}).

\bibitem{Fuechsle_NN_2012}
\bibinfo{author}{Fuechsle, M.} \emph{et~al.}
\newblock \bibinfo{title}{A single-atom transistor}.
\newblock \emph{\bibinfo{journal}{Nature Nanotechnology}}
  \textbf{\bibinfo{volume}{7}}, \bibinfo{pages}{242} (\bibinfo{year}{2012}).

\bibitem{Salfi_PRX_2018}
\bibinfo{author}{Salfi, J.} \emph{et~al.}
\newblock \bibinfo{title}{Valley filtering in spatial maps of coupling between
  silicon donors and quantum dots}.
\newblock \emph{\bibinfo{journal}{Phys. Rev. X}} \textbf{\bibinfo{volume}{8}},
  \bibinfo{pages}{031049} (\bibinfo{year}{2018}).

\bibitem{Pla_Nature_2012}
\bibinfo{author}{Pla, J.} \emph{et~al.}
\newblock \bibinfo{title}{A single-atom electron spin qubit in silicon}.
\newblock \emph{\bibinfo{journal}{Nature}} \textbf{\bibinfo{volume}{489}},
  \bibinfo{pages}{541} (\bibinfo{year}{2012}).

\bibitem{Morello_Nature_2010}
\bibinfo{author}{Morello, A.} \emph{et~al.}
\newblock \bibinfo{title}{Single-shot readout of an electron spin in silicon}.
\newblock \emph{\bibinfo{journal}{Nature}} \textbf{\bibinfo{volume}{467}},
  \bibinfo{pages}{687} (\bibinfo{year}{2010}).

\bibitem{Tyryshkin_Nat_Mat_2012}
\bibinfo{author}{Tyryshkin, A.} \emph{et~al.}
\newblock \bibinfo{title}{Electron spin coherence exceeding seconds in
  high-purity silicon}.
\newblock \emph{\bibinfo{journal}{Nature Materials}}
  \textbf{\bibinfo{volume}{11}}, \bibinfo{pages}{143} (\bibinfo{year}{2012}).

\bibitem{Saeedi_Science_2013}
\bibinfo{author}{Saeedi, K.} \emph{et~al.}
\newblock \bibinfo{title}{Room-temperature quantum bit storage exceeding 39
  minutes using ionized donors in silicon-28}.
\newblock \emph{\bibinfo{journal}{Science}} \textbf{\bibinfo{volume}{342}},
  \bibinfo{pages}{830} (\bibinfo{year}{2013}).

\bibitem{Pica_PRB_2016}
\bibinfo{author}{Pica, G.}, \bibinfo{author}{Lovett, B.~W.},
  \bibinfo{author}{Bhatt, R.~N.}, \bibinfo{author}{Schenkel, T.} \&
  \bibinfo{author}{Lyon, S.~A.}
\newblock \bibinfo{title}{Surface code architecture for donors and dots in
  silicon with imprecise and nonuniform qubit couplings}.
\newblock \emph{\bibinfo{journal}{Phys. Rev. B}} \textbf{\bibinfo{volume}{93}},
  \bibinfo{pages}{035306} (\bibinfo{year}{2016}).

\bibitem{Wang_NatureSR_2016}
\bibinfo{author}{Wang, Y.}, \bibinfo{author}{Chen, C.-Y.},
  \bibinfo{author}{Klimeck, G.}, \bibinfo{author}{Simmons, M.} \&
  \bibinfo{author}{Rahman, R.}
\newblock \bibinfo{title}{Characterizing si:p quantum dot qubits with spin
  resonance techniques}.
\newblock \emph{\bibinfo{journal}{Scientific Reports}}
  \textbf{\bibinfo{volume}{6}}, \bibinfo{pages}{31830} (\bibinfo{year}{2016}).

\bibitem{Testlin_PRA_2007}
\bibinfo{author}{Testolin, M.}, \bibinfo{author}{Hill, C.},
  \bibinfo{author}{Wellard, C.} \& \bibinfo{author}{Hollenberg, L.}
\newblock \bibinfo{title}{A precise cnot gate in the presence of large
  fabrication induced variations of the exchange interaction strength}.
\newblock \emph{\bibinfo{journal}{Phys. Rev. A}} \textbf{\bibinfo{volume}{76}},
  \bibinfo{pages}{012302} (\bibinfo{year}{2007}).

\bibitem{Hill_PRL_2007}
\bibinfo{author}{Hill, C.}
\newblock \bibinfo{title}{Robust controlled-not gates from almost any
  interaction}.
\newblock \emph{\bibinfo{journal}{Phys. Rev. Lett.}}
  \textbf{\bibinfo{volume}{98}}, \bibinfo{pages}{180501}
  (\bibinfo{year}{2007}).

\bibitem{Butler_Nature_2018}
\bibinfo{author}{Butler, K.}, \bibinfo{author}{Davies, D.},
  \bibinfo{author}{Cartwright, H.}, \bibinfo{author}{Isayev, O.} \&
  \bibinfo{author}{Walsh, A.}
\newblock \bibinfo{title}{Machine learning for molecular and materials
  science}.
\newblock \emph{\bibinfo{journal}{Nature}} \textbf{\bibinfo{volume}{559}},
  \bibinfo{pages}{547} (\bibinfo{year}{2018}).

\bibitem{Luna_Nature_2017}
\bibinfo{author}{Luna, P.}, \bibinfo{author}{Wei, J.}, \bibinfo{author}{Bengio,
  Y.}, \bibinfo{author}{Aspuru-Guzik, A.} \& \bibinfo{author}{Sargent, E.}
\newblock \bibinfo{title}{Use machine learning to find energy materials}.
\newblock \emph{\bibinfo{journal}{Nature}} \textbf{\bibinfo{volume}{552}},
  \bibinfo{pages}{23} (\bibinfo{year}{2017}).

\bibitem{Libbrecht_NatureRG_2015}
\bibinfo{author}{Libbrecht, M.} \& \bibinfo{author}{Noble, W.}
\newblock \bibinfo{title}{Machine learning applications in genetics and
  genomics}.
\newblock \emph{\bibinfo{journal}{Nature Reviews Genetics}}
  \textbf{\bibinfo{volume}{16}}, \bibinfo{pages}{321} (\bibinfo{year}{2015}).

\bibitem{Murphy_NatureCB_2011}
\bibinfo{author}{Murphy, R.}
\newblock \bibinfo{title}{An active role for machine learning in drug
  development}.
\newblock \emph{\bibinfo{journal}{Nature Chemical Biology}}
  \textbf{\bibinfo{volume}{7}}, \bibinfo{pages}{327} (\bibinfo{year}{2011}).

\bibitem{Heureux_IEEE_2017}
\bibinfo{author}{Heureux, A.}, \bibinfo{author}{Grolinger, K.},
  \bibinfo{author}{Elyamany, H.} \& \bibinfo{author}{Capretz, M.}
\newblock \bibinfo{title}{Machine learning with big data: Challenges and
  approaches}.
\newblock \emph{\bibinfo{journal}{IEEE Access}} \textbf{\bibinfo{volume}{5}},
  \bibinfo{pages}{7776} (\bibinfo{year}{2017}).

\bibitem{Rashidi_ACSN_2018}
\bibinfo{author}{Rashidi, M.} \& \bibinfo{author}{Wolkow, R.}
\newblock \bibinfo{title}{Autonomous scanning probe microscopy in situ tip
  conditioning through machine learning}.
\newblock \emph{\bibinfo{journal}{ACS Nano}} \textbf{\bibinfo{volume}{12 (6)}},
  \bibinfo{pages}{5185} (\bibinfo{year}{2018}).

\bibitem{Rashidi_arXiv_2019}
\bibinfo{author}{Rashidi, M.} \emph{et~al.}
\newblock \bibinfo{title}{Autonomous atomic scale manufacturing through machine
  learning}.
\newblock \emph{\bibinfo{journal}{arXiv:1902.08818}}  (\bibinfo{year}{2019}).

\bibitem{bishop2006pattern}
\bibinfo{author}{Bishop, C.~M.}
\newblock \emph{\bibinfo{title}{Pattern recognition and machine learning}}
  (\bibinfo{publisher}{springer}, \bibinfo{year}{2006}).

\bibitem{Salfi_NatMat_2014}
\bibinfo{author}{Salfi, J.} \emph{et~al.}
\newblock \bibinfo{title}{Spatially resolving valley quantum interference of a
  donor in silicon}.
\newblock \emph{\bibinfo{journal}{Nature Materials}}
  \textbf{\bibinfo{volume}{13}}, \bibinfo{pages}{605} (\bibinfo{year}{2014}).

\bibitem{Usman_NN_2016}
\bibinfo{author}{Usman, M.} \emph{et~al.}
\newblock \bibinfo{title}{Spatial metrology of dopants in silicon with exact
  lattice site precision}.
\newblock \emph{\bibinfo{journal}{Nature Nanotechnology}}
  \textbf{\bibinfo{volume}{11}}, \bibinfo{pages}{763} (\bibinfo{year}{2016}).

\bibitem{Sinthiptharakoon_JPCM_2014}
\bibinfo{author}{Sinthiptharakoon, K.} \emph{et~al.}
\newblock \emph{\bibinfo{journal}{J. Phys.: Condens. Matter}}
  \textbf{\bibinfo{volume}{26}}, \bibinfo{pages}{012001}
  (\bibinfo{year}{2014}).

\bibitem{Garleff_PRB_2008}
\bibinfo{author}{Garleff, J.~K.} \emph{et~al.}
\newblock \emph{\bibinfo{journal}{Phys. Rev. B}} \textbf{\bibinfo{volume}{78}},
  \bibinfo{pages}{075313} (\bibinfo{year}{2008}).

\bibitem{Ishida_Nanoscale_2015}
\bibinfo{author}{Ishida, N.} \emph{et~al.}
\newblock \bibinfo{title}{Direct visualization of the n impurity state in
  dilute ganas using scanning tunneling microscopy}.
\newblock \emph{\bibinfo{journal}{Nanoscale}} \textbf{\bibinfo{volume}{7}},
  \bibinfo{pages}{16773} (\bibinfo{year}{2015}).

\bibitem{Plantenga_PRB_2007}
\bibinfo{author}{Plantenga, R.} \emph{et~al.}
\newblock \bibinfo{title}{Spatially resolved electronic structure of an
  isovalent nitrogen center in gaas}.
\newblock \emph{\bibinfo{journal}{Phys. Rev. B}} \textbf{\bibinfo{volume}{96}},
  \bibinfo{pages}{155210} (\bibinfo{year}{2017}).

\bibitem{Krammel_PRM_2017}
\bibinfo{author}{Krammel, C.} \emph{et~al.}
\newblock \bibinfo{title}{Incorporation of bi atoms in inp studied at the
  atomic scale by cross-sectional scanning tunneling microscopy}.
\newblock \emph{\bibinfo{journal}{Phys. Rev. Materials}}
  \textbf{\bibinfo{volume}{1}}, \bibinfo{pages}{034606} (\bibinfo{year}{2017}).

\bibitem{Brazdova_PRB_2017}
\bibinfo{author}{Brazdova, V.} \emph{et~al.}
\newblock \bibinfo{title}{Exact location of dopants below the si(001):h surface
  from scanning tunneling microscopy and density functional theory}.
\newblock \emph{\bibinfo{journal}{Phys. Rev. B}} \textbf{\bibinfo{volume}{95}},
  \bibinfo{pages}{075408} (\bibinfo{year}{2017}).

\bibitem{Usman_Nanoscale_2017}
\bibinfo{author}{Usman, M.}, \bibinfo{author}{Voisin, B.},
  \bibinfo{author}{Salfi, J.}, \bibinfo{author}{Rogge, S.} \&
  \bibinfo{author}{Hollenberg, L. C.~L.}
\newblock \bibinfo{title}{Towards visualisation of central-cell-effects in
  scanning tunnelling microscope images of subsurface dopant qubits in
  silicon}.
\newblock \emph{\bibinfo{journal}{Nanoscale}} \textbf{\bibinfo{volume}{9}},
  \bibinfo{pages}{17013} (\bibinfo{year}{2017}).

\bibitem{Wang_NQI_2016}
\bibinfo{author}{Wang, Y.} \emph{et~al.}
\newblock \bibinfo{title}{Highly tunable exchange in donor qubits in silicon}.
\newblock \emph{\bibinfo{journal}{npj Quantum Information}}
  \textbf{\bibinfo{volume}{2}}, \bibinfo{pages}{16008} (\bibinfo{year}{2016}).

\bibitem{Pakkiam_PRX_2018}
\bibinfo{author}{Pakkiam, P.} \emph{et~al.}
\newblock \bibinfo{title}{Single-shot single-gate rf spin readout in silicon}.
\newblock \emph{\bibinfo{journal}{Phys. Rev. X}} \textbf{\bibinfo{volume}{8}},
  \bibinfo{pages}{041032} (\bibinfo{year}{2018}).

\bibitem{Usman_JPCM_2015}
\bibinfo{author}{Usman, M.} \emph{et~al.}
\newblock \bibinfo{title}{Donor hyperfine stark shift and the role of
  central-cell corrections in tight-binding theory}.
\newblock \emph{\bibinfo{journal}{J. Phys.: Cond. Matt.}}
  \textbf{\bibinfo{volume}{27}}, \bibinfo{pages}{154207}
  (\bibinfo{year}{2015}).

\bibitem{Bardeen_PRL_1961}
\bibinfo{author}{Bardeen, J.}
\newblock \bibinfo{title}{Tunnelling from a many-particle point of view}.
\newblock \emph{\bibinfo{journal}{Phys. Rev. Lett.}}
  \textbf{\bibinfo{volume}{6}}, \bibinfo{pages}{57} (\bibinfo{year}{1961}).

\bibitem{Chen_PRB_1990}
\bibinfo{author}{Chen, C.~J.}
\newblock \bibinfo{title}{Tunneling matrix elements in three-dimensional space:
  The derivative rule and the sum rule}.
\newblock \emph{\bibinfo{journal}{Phys. Rev. B}} \textbf{\bibinfo{volume}{42}},
  \bibinfo{pages}{8841} (\bibinfo{year}{1990-I}).

\bibitem{Kingma_arxiv_2017}
\bibinfo{author}{Kingma, D.} \& \bibinfo{author}{Ba, J.}
\newblock \bibinfo{title}{Adam: A method for stochastic optimization}.
\newblock \emph{\bibinfo{journal}{arXiv:1412.6980v9}}  (\bibinfo{year}{2017}).

\bibitem{chollet2015keras}
\bibinfo{author}{Chollet, F.} \emph{et~al.}
\newblock \bibinfo{title}{Keras}.
\newblock \bibinfo{howpublished}{\url{https://github.com/fchollet/keras}}
  (\bibinfo{year}{2015}).

\bibitem{tensorflow2015-whitepaper}
\bibinfo{author}{Abadi, M.} \emph{et~al.}
\newblock \bibinfo{title}{{TensorFlow}: Large-scale machine learning on
  heterogeneous systems} (\bibinfo{year}{2015}).
\newblock \urlprefix\url{https://www.tensorflow.org/}.
\newblock \bibinfo{note}{Software available from tensorflow.org}.

\bibitem{Boykin_PRB_2004}
\bibinfo{author}{Boykin, T.~B.}, \bibinfo{author}{Klimeck, G.} \&
  \bibinfo{author}{Oyafuso, F.}
\newblock \bibinfo{title}{Valence band effective-mass expressions in the
  sp3d5s* empirical tight-binding model applied to a si and ge
  parametrization}.
\newblock \emph{\bibinfo{journal}{Phys. Rev. B}} \textbf{\bibinfo{volume}{69}},
  \bibinfo{pages}{115201} (\bibinfo{year}{2004}).

\bibitem{Nara_JPSJ_1965}
\bibinfo{author}{Nara, H.}
\newblock \bibinfo{title}{Screened impurity potential in si}.
\newblock \emph{\bibinfo{journal}{J. Phys. Soc. Jap.}}
  \textbf{\bibinfo{volume}{20}}, \bibinfo{pages}{778} (\bibinfo{year}{1965}).

\bibitem{Overhof_PRL_2004}
\bibinfo{author}{Overhof, H.} \& \bibinfo{author}{Gerstmann, U.}
\newblock \bibinfo{title}{Ab initio calculation of hyperfine and superhyperfine
  interactions for shallow donors in semiconductors}.
\newblock \emph{\bibinfo{journal}{Phys. Rev. Lett.}}
  \textbf{\bibinfo{volume}{92}}, \bibinfo{pages}{087602}
  (\bibinfo{year}{2004}).

\bibitem{Usman_PRB_2015}
\bibinfo{author}{Usman, M.} \emph{et~al.}
\newblock \bibinfo{title}{Strain and electric field control of hyperfine
  interactions for donor spin qubits in silicon}.
\newblock \emph{\bibinfo{journal}{Phys. Rev. B}} \textbf{\bibinfo{volume}{91}},
  \bibinfo{pages}{245209} (\bibinfo{year}{2015}).

\bibitem{Craig_SS_1990}
\bibinfo{author}{Craig, B.~I.} \& \bibinfo{author}{Smith, P.~V.}
\newblock \bibinfo{title}{The structure of the si(100)2 × 1: H surface}.
\newblock \emph{\bibinfo{journal}{Surface Science}}
  \textbf{\bibinfo{volume}{226}}, \bibinfo{pages}{L55} (\bibinfo{year}{1990}).

\bibitem{Lee_PRB_2004}
\bibinfo{author}{Lee, S.}, \bibinfo{author}{Oyafuso, F.}, \bibinfo{author}{von
  Allmen, P.} \& \bibinfo{author}{Klimeck, G.}
\newblock \bibinfo{title}{Boundary conditions for the electronic structure of
  finite-extent embedded semiconductor nanostructures}.
\newblock \emph{\bibinfo{journal}{Phys. Rev. B}} \textbf{\bibinfo{volume}{69}},
  \bibinfo{pages}{045316} (\bibinfo{year}{2004}).

\bibitem{Klimeck_2}
\bibinfo{author}{Klimeck, G.} \emph{et~al.}
\newblock \emph{\bibinfo{journal}{IEEE Trans. Elect. Dev.}}
  \textbf{\bibinfo{volume}{54}}, \bibinfo{pages}{2090} (\bibinfo{year}{2007}).

\bibitem{Ahmed_Enc_2009}
\bibinfo{author}{Ahmed, S.} \emph{et~al.}
\newblock \bibinfo{title}{Multimillion atom simulations with nemo 3-d}.
\newblock \emph{\bibinfo{journal}{Springer Encyclopedia of Complexity and
  Systems Science ed R A Meyers (Heidelberg: Springer)}}
  \bibinfo{pages}{p5745--83} (\bibinfo{year}{2009}).

\bibitem{Slater_PR_1954}
\bibinfo{author}{Slater, J.~C.} \& \bibinfo{author}{Koster, G.~F.}
\newblock \bibinfo{title}{Simplified lcao method for the periodic potential
  problem}.
\newblock \emph{\bibinfo{journal}{Phys. Rev.}} \textbf{\bibinfo{volume}{94}},
  \bibinfo{pages}{1498} (\bibinfo{year}{1954}).

\end{thebibliography}

\end{document}